\documentclass[12pt]{article}
\pdfoutput=1

\usepackage[auth-sc]{authblk}
\usepackage{booktabs}

\title{\MakeUppercase{Why Do the Elderly Save? Using Health Shocks to Uncover Bequests Motives}}

\author[1]{Tetsuya Kaji}
\author[2]{Elena Manresa}
\affil[1]{University of Chicago}
\affil[2]{Princeton University}

\date{\normalsize\today}

\usepackage[margin=1.25in]{geometry}
\usepackage[onehalfspacing,nodisplayskipstretch]{setspace} 

\usepackage{amsthm,amsmath,amssymb,calc}
\usepackage{thmtools}
\usepackage{mathrsfs}
\usepackage{mathtools}
\usepackage{bm}
\usepackage{bbm}

\usepackage{titling}
\usepackage[tiny]{titlesec}
\usepackage{chngcntr}

\usepackage[authoryear]{natbib}
\usepackage{breakcites}
\RequirePackage[colorlinks,citecolor=blue,urlcolor=blue]{hyperref}
\RequirePackage{cleveref}
\usepackage{hyphenat} 

\usepackage{chapterbib} 
\usepackage{enumerate}
\usepackage{rotating}
\usepackage{xcolor}

\usepackage{graphicx,epstopdf,epsfig}
\usepackage{caption,subcaption}
\usepackage{pgfplots}

\usepackage[T1]{fontenc}
\usepackage{textcomp}
\usepackage{lmodern}

\pretitle{\vspace{-5em}\begin{center}\fontsize{13}{15}\selectfont}
\posttitle{\par\end{center}}
\preauthor{\vspace{-1em}\begin{center}
	\normalsize \scshape \lineskip 0.5em%
	\begin{tabular}[t]{c}}
\postauthor{\end{tabular}\par\end{center}}
\predate{\begin{center}\large}
\postdate{\par\end{center}\vspace{-1em}}

\makeatletter
\newcommand\new@setfontsize[3]{%
    \ifx \protect \@typeset@protect \let \@currsize #1\fi \fontsize {#2}{#3}\selectfont
}
\let\orig@setfontsize\@setfontsize
\let\orig@cases\cases
\let\endorig@cases\endcases

\renewenvironment{cases}{%
    \let\@setfontsize\new@setfontsize
    \setstretch{\setspace@singlespace}%
    \let\setfontsize\orig@setfontsize
    \orig@cases
}{%
    \endorig@cases
}
\makeatother

\titleformat{\section}{\normalfont\centering}{\thesection}{1em}{\MakeUppercase}
\titleformat*{\subsection}{\itshape\centering}

\theoremstyle{plain}

\newtheorem*{thm*}{Theorem}

\theoremstyle{definition}
\newtheorem*{defn}{Definition}

\newtheorem*{exa*}{Example}
\newtheorem*{alg}{Algorithm}

\theoremstyle{remark}

\renewcommand\thmcontinues[1]{continued}

\crefname{thm}{Theorem}{Theorems}
\crefname{prop}{Proposition}{Propositions}
\crefname{lem}{Lemma}{Lemmas}
\crefname{coro}{Corollary}{Corollaries}
\crefname{add}{Addendum}{Addendums}
\crefname{asm}{Assumption}{Assumptions}
\crefname{alg}{Algorithm}{Algorithms}
\crefname{proc}{Procedure}{Procedures}
\crefname{exa}{Example}{Examples}
\crefname{section}{Section}{Sections}
\crefname{subsection}{Section}{Sections}
\crefname{appendix}{Appendix}{Appendices}

\crefrangelabelformat{exa}{#3#1#4--#5\crefstripprefix{#1}{#2}#6}

\def\argmin{\mathop{\arg\min}}		

\makeatletter
\def\blfootnote{\gdef\@thefnmark{}\@footnotetext}
\makeatother

\allowdisplaybreaks[3]

\counterwithin{subsection}{section}

\begin{document}


\defcitealias{dfj}{DFJ}

\maketitle

\begin{abstract}
We revisit the saving behavior of elderly singles using an adversarial structural estimation framework by \cite{kmp2023adversarial}. The method bridges the simulated method of moments (SMM) and maximum-likelihood estimation by embedding a flexible discriminator—implemented as a neural network—that adaptively selects the most informative features of the data. Applying this approach to the model of \cite{dfj} with AHEAD data, we show that including gender and health histories in the discriminator improves identification and precision of bequests motives. The resulting estimates reveal that bequest motives explain between $13\%$ and $19\%$ percent of late-life savings across all permanent-income quintiles, not only among the rich. The adversarial estimator precisely disentangles bequest motives from precautionary savings motives. These findings suggest that heterogeneity in health-related survival expectations is another important source of identifying variation to distinguishing bequest and precautionary saving motives.
\bigskip

\vspace{4pt}

\vspace{4pt}
\end{abstract}

\blfootnote{We thank Mariacristina De Nardi and John Jones for sharing the data and codes and for very helpful discussion. Aziza Kurvonova (ak6330@nyu.edu) provided invaluable computation assistance. We also thank the HPC Greene team at NYU, including Shenglong Wang and Sergey Samsonau. We gratefully acknowledge the support of the NSF by means of the Grant SES\hyp{}1824304 and the Richard N.\ Rosett Faculty Fellowship and the Liew Family Faculty Fellowship at the University of Chicago Booth School of Business. Any errors are our own.}


\section{Introduction}\label{sec:1}

The elderly save for a variety of reasons: uncertainty about survival, the desire to leave bequests, and rising medical expenses as they age. Distinguishing among these motives is essential for understanding the mechanisms behind wealth decumulation and for evaluating the impact of social insurance programs such as Medicaid and Medicare. Despite extensive research, the relative importance of precautionary and bequest motives remains difficult to identify empirically, as both lead to slow asset drawdown late in life.

The risks faced by the elderly are highly heterogeneous, varying with gender, age, health status, and permanent income. These differences imply substantial heterogeneity in saving behavior: women and the rich tend to live longer than men and the poor, which mechanically affects the incentives to accumulate or decumulate wealth. Failing to account for these differences can bias inference about preferences, making the rich appear more patient or more altruistic than they truly are. Yet incorporating this heterogeneity into standard estimation methods such as the simulated method of moments (SMM) typically leads to large sampling variability. Selecting moments that capture all relevant sources of variation is difficult, and the precision of the estimates deteriorates quickly as the number of moments grows. This motivates the use of a more flexible estimation approach that can exploit complex heterogeneity without loss of efficiency.

We address this challenge by combining the life-cycle model of the seminal paper \cite{dfj} with the adversarial structural estimation framework \cite{kmp2023adversarial}. The adversarial method searches for parameter values that make simulated data generated by the model statistically indistinguishable from the observed data. It embeds a flexible discriminator, implemented as a neural network, that adaptively selects the most informative features of the data. When the discriminator is logistic, the estimator coincides with optimally weighted SMM; when it approximates the likelihood ratio, it attains the efficiency of maximum likelihood estimation. 

In order to provide evidence of the performance of the method, we conduct a large Monte Carlo exercise calibrated to the \cite{dfj} model. Across multiple specifications, the adversarial estimator recovers the true structural parameters with negligible bias and significant precision. In particular, precision improves substantially when the discriminator incorporates information on health and gender, confirming that these variables contain valuable identifying variation. The loss function behaves quadratically around the true parameters, indicating that the estimator is well approximated by a Gaussian distribution and that inference based on bootstrap or asymptotic arguments is reliable. These results provide strong evidence that the adversarial framework is accurate, computationally stable, and useful in this context.

Using data from the Assets and Health Dynamics Among the Oldest Old (AHEAD) survey, we re-estimate the parameters governing risk aversion and bequest motives for single retirees aged seventy and older. Including gender and health histories in the discriminator provides additional identifying variation and improves precision. The resulting estimates imply that bequest motives explain a larger proportion of elderly savings across all permanent-income quintiles, not only among the rich. The implied asset floor is around the twenty-second percentile of the wealth distribution, suggesting that many elderly individuals derive utility from leaving bequests even at moderate levels of wealth.

Our paper contributes to a long line of research on the motives behind late-life saving. The early literature, culminating in \cite{dfj}, emphasizes medical-expense risk as the primary driver of wealth decumulation among the elderly. Related studies such as \cite{k2007} stress the role of social insurance and estate taxation, while \cite{kopczuk2007leave} model explicit bequest motives but conclude that they are economically significant only for the wealthy. These papers established the benchmark view that precautionary savings against medical and longevity risk largely account for the observed slow drawdown of assets at older ages.

A new wave of work has re-examined this conclusion by introducing richer heterogeneity and new sources of identifying variation to separate bequest from precautionary motives. As a first example, \cite{lockwood2018incidental} exploits variation in annuitization choices to infer altruistic preferences, showing that bequest motives explain roughly fifteen to twenty-five percent of retirement wealth. \cite{ameriks2020long} combine survey data with administrative records to measure subjective bequest intentions directly and find comparable magnitudes. \cite{de2025couples} extend the \cite{dfj} model to couples and singles, incorporating gender, health, and marital heterogeneity, and conclude that residual bequest motives—both intended and incidental—account for ten to twenty percent of late-life assets. These studies illustrate that adding additional variation—through annuitization behavior, self-reported intentions, or detailed survival heterogeneity—sharply improves identification of altruistic preferences.

Our contribution follows this tradition by using health shocks and gender differences in survival expectations as new sources of identifying variation within a structural framework. The adversarial estimation approach allows these dimensions of heterogeneity to be exploited flexibly and efficiently, yielding estimates that fall within the ten to twenty percent range but demonstrate that bequest motives are pervasive across the entire permanent-income distribution rather than concentrated among the wealthy.

We contribute to this literature in three ways. First, methodologically, we adapt the adversarial estimation framework of \cite{kmp2023adversarial} to a standard life-cycle model, offering a bridge between SMM and MLE. Second, empirically, we show that exploiting gender- and health-driven variation in survival expectations yields much sharper identification of bequest motives. Third, we provide new quantitative evidence that bequest motives are economically relevant throughout the wealth distribution, not just at the top.

The rest of the paper is organized as follows:
\cref{sec:2} introduces the model, 
\cref{sec:data} describes the data, \cref{sec:id} provides an identification discussion, \cref{sec:est} explains the estimator, with details on implementation, \cref{sec:MC} presents results on an extensive Monte Carlo exercise to provide evidence of the validity of our estimation framework. Finally, 
\cref{sec:res} covers our findings and \cref{sec:conc} concludes. 

\section{A Model of Savings for the Elderly}\label{sec:2}

We follow \citetalias{dfj} for modeling savings behavior. As them, we focus on the behavior of single, retired individuals of age 70 and older. We explain the model below for reading convenience, but there is no changes realtive to DFJ. 
In each period, a surviving single retired agent receives utility $u(c)$ from consumption $c$ and, if they die in that period, additional utility $\phi(e)$ from leaving estate $e$, where
\[
	u(c)\vcentcolon=\frac{c^{1-\nu}}{1-\nu}, \qquad\quad
	\phi(e)\vcentcolon=\vartheta\frac{(e+k)^{1-\nu}}{1-\nu},
\]
and $\nu$ is the relative risk aversion and $\vartheta$ and $k$ are the intensity and curvature of the bequest motive.
Each individual is associated with gender $g$ and permanent income $I$, and carries six state variables: age $t$, asset $a_t$, nonasset income $y_t$, health status $h_t$, medical expense shock $\zeta_t$, and survival $s_t$.
Health and survival are binary, where $h_t=1$ means they are healthy at age $t$, and $s_t=1$ they survive to the next period.

They face three channels of uncertainty: health, survival, and medical expenses.
Heath and survival evolve as Markov chains. We denote
\[
	\pi_{H}(g,h_t,I,t)\vcentcolon=\Pr(h_{t+1}=1\mid g,h_t,I,t), \quad
	\pi_{S}(g,h_t,I,t)\vcentcolon=\Pr(s_{t+1}=1\mid g,h_t,I,t).
\]
The medical expenses they incur are given by
\[
	\log m_{t}=m(g,h_t,I,t)+\sigma(g,h_t,I,t)\times\psi_{t},
\]
where $m$ and $\sigma$ are deterministic functions, 
$\psi_{t}=\zeta_{t}+\xi_{t}$, $\xi_{t}\sim N(0,\sigma_{\xi}^{2})$, $\zeta_{t}=\rho\zeta_{t-1}+\epsilon_{t}$, and $\epsilon_{t}\sim N(0,\sigma_{\epsilon}^{2})$.
The nonasset income evolves deterministically as $y_t=y(g,I,t)$.
The asset evolves as
\[
	a_{t+1}=a_{t}+y_{n}(ra_{t}+y_{t},\tau)+b_{t}-m_{t}-c_t,
\]
where $b_{t}\geq0$ is the {\em government transfer}, $r$ the {\em risk\hyp{}free pretax rate of return}, $y_{n}(\cdot,\tau)$ the {\em posttax income}, and $\tau$ the {\em tax structure}.
The agent faces a borrowing constraint $a_t\geq0$ while social insurance guarantees minimum consumption $c_t\geq\underline{c}$; government transfer $b_t$ is positive only when both constraints cannot be satisfied without it.

The timing in each period is given as follows. Heath $h_t$ and medical expenses $m_t$ realize; then the individual chooses consumption $c_t$; then survival $s_t$ realizes; if $s_t=0$, they leave the remaining assets as bequest; if $s_t=1$, move on to the next period.


Denoting the {\em cash\hyp{}on\hyp{}hand} by $x_t\vcentcolon=c_t+a_{t+1}$, the agent's Bellman equation is
\[
	V_{t}(x,g,h,I,\zeta)=\max_{c,x'} \, u(c,h)+\beta[s\mathbb{E}_tV_{t+1}(x',g,h',I,\zeta')+(1-s)\phi(e)]
\]
subject to $x'=(x-c)+y_{n}(r(x-c)+y',\tau)+b'-m'$, $e=(x-c)-\max\{0,\tilde{\tau}(x-c-\tilde{x})\}$, and $x\geq c\geq\underline{c}$.
The first constraint is the budget constraint; the second the bequest (taxed at rate $\tilde{\tau}$ with deduction $\tilde{x}$); the last are the borrowing and consumption constraints.

We also look at two transformations: the {\em marginal propensity to consume at the moment of death} $\text{MPC}\vcentcolon=(1+r)/(1+r+[\beta\vartheta(1+r)]^{1/\nu})$ and the {\em implied asset floor} $\underline{a}\vcentcolon=k/[\beta\vartheta(1+r)]^{1/\nu}$ above which individuals get utility from bequeathing.%
\footnote{The {\em marginal propensity to bequeath (MPB)} is defined by $1-\text{MPC}$.}

\section{Data}\label{sec:data}

We use the same data as \citetalias{dfj}, taken from {\em Assets and Health Dynamics Among the Oldest Old (AHEAD)}.
%
%
The sample consists of non\hyp{}institutionalized individuals of age 70 and older in 1994. It contains 8,222 individuals in 6,047 households (3,872 singles and 2,175 couples).
The survey took place biyearly from 1994 to 2006.
We focus on 3,259 single retired individuals, 592 of which are men and 2,667 women.%
\footnote{Single individuals are those who were neither married nor cohabiting at any point in the analysis.}
Of those, 884 were alive in 2006.
We drop the first survey in 1994 for reliability, following \citetalias{dfj}.

The survey collects information on age at $t$, financial wealth $a_t$, nonasset income $y_t$, medical expenses $m_t$, and health status $h_t$. Financial wealth includes real estate, autos, several other liquid assets, retirement accounts, etc. Nonasset income includes social security benefits, veteran's benefits, and other benefits. Medical expenses are total out\hyp{}of\hyp{}pocket spending; the average yearly expenses are \$3,700 with standard deviation \$13,400. The permanent income is not observed, but we use as a proxy the ranking of individual average income over time.
The health status is a binary variable indicating whether the individual perceives themself as healthy.

\section{Identification Discussion}\label{sec:id}

Disentangling the bequest motive from medical expenditure risk is a challenging task. As the bequest is modeled as a luxury good, we expect that identifying power comes mainly from variation of wealthy individuals. Meanwhile, wealthy individuals are also the ones with the longest life expectancy, ane hence with the most incentives to save for medical expenses.

Indeed, \citetalias{dfj} document that the medical expenditure for the rich skyrockets after age 95, reaching \$15,000 by age 100.
However, if their health condition diminishes their life expectancy, those with shorter horizons would face much less incentive to save for the coming medical expenses while a stronger incentive to save for bequests.

We now investigate if information on health status could be helpful to identify savings' bequest motives separately from medical expenditure savings's motives.\footnote{See for example \citep{kopecky2010impact} and references therein on the role of health to identify bequests.}. We now provide suggestive evidence that it is the case in our data.

\begin{figure}[htbp]
\caption{Profiles by gender and health}
\centering
\begin{subfigure}[t]{0.45\textwidth}
\centering \includegraphics[width=\textwidth]{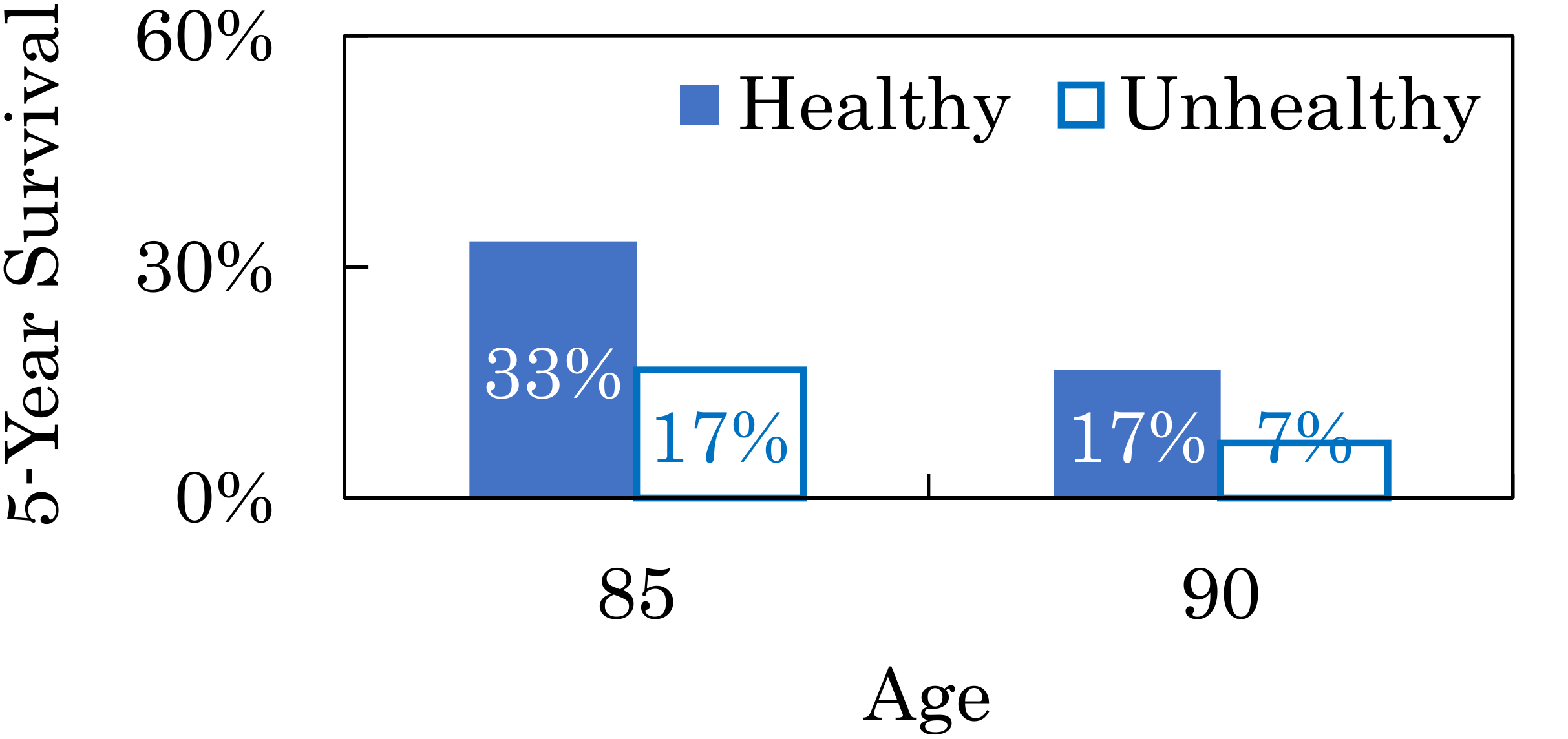}
\caption{Men's five\hyp{}year survival rates.} \label{fig:1g}
\end{subfigure}
\qquad
\begin{subfigure}[t]{0.45\textwidth}
\centering \includegraphics[width=\textwidth]{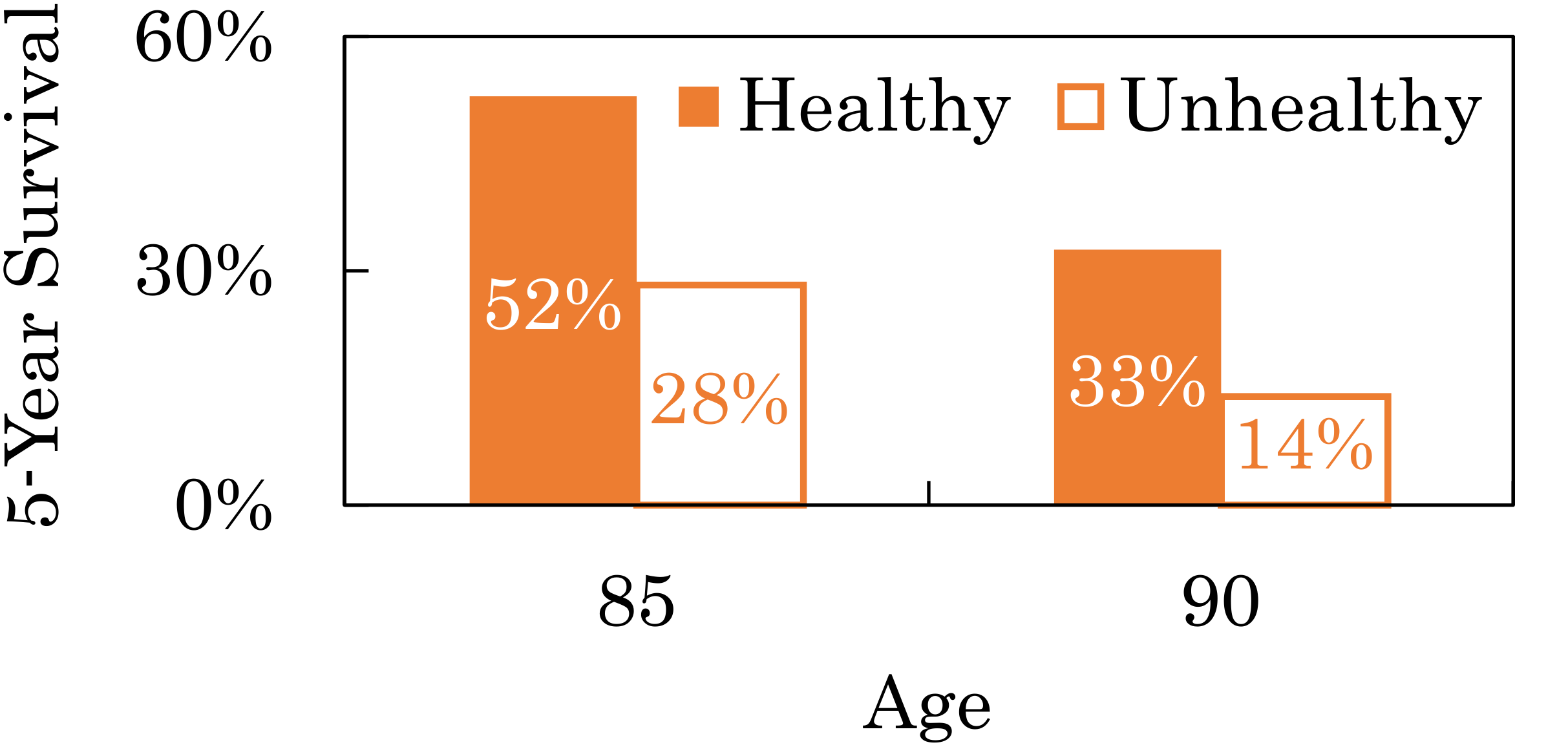}
\caption{Women's five\hyp{}year survival rates.} \label{fig:1h}
\end{subfigure}
\\\vspace{6pt}
\begin{subfigure}[t]{0.45\textwidth}
\centering \includegraphics[width=\textwidth]{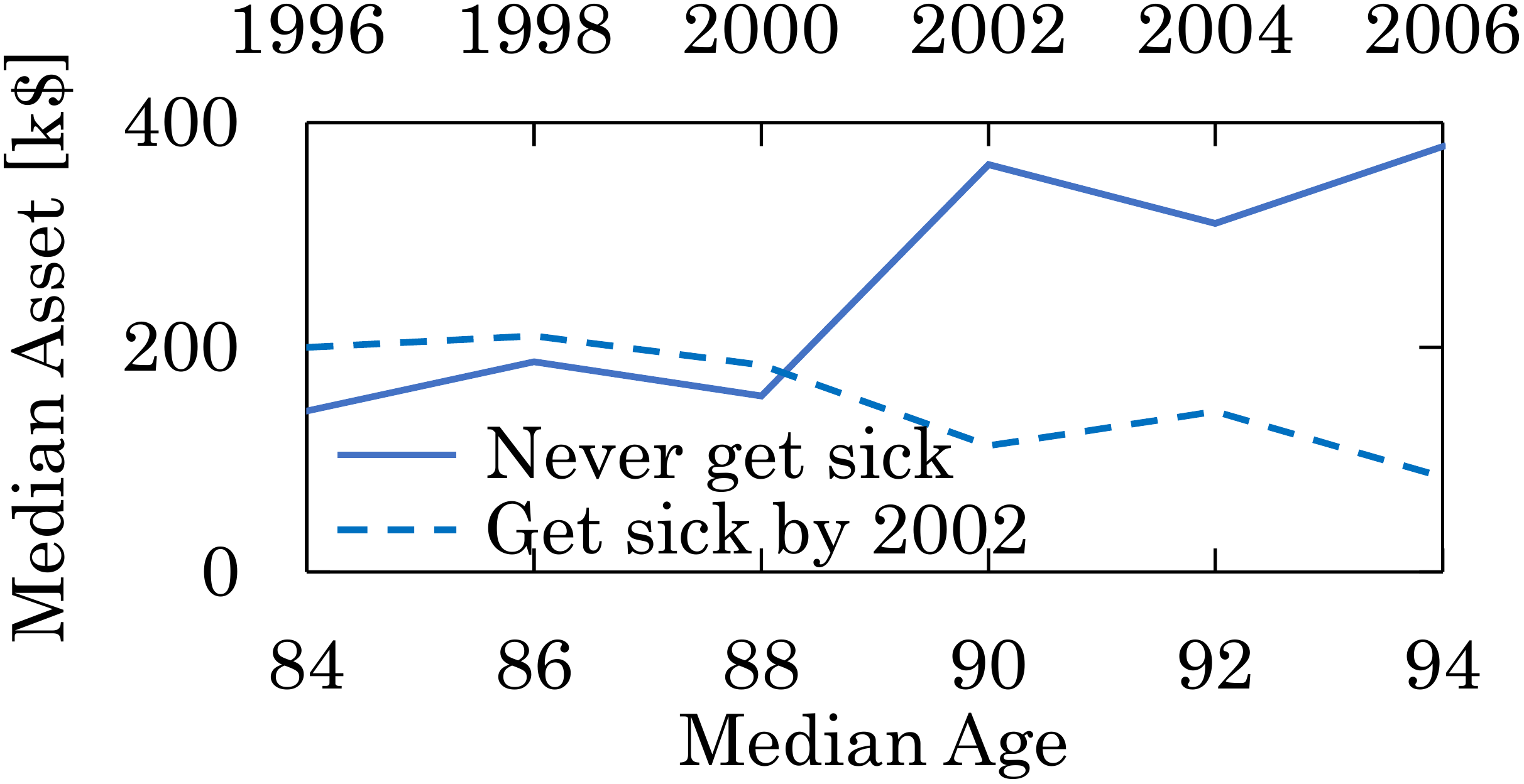}
\caption{Men's asset.} \label{fig:1a}
\end{subfigure}
\qquad
\begin{subfigure}[t]{0.45\textwidth}
\centering \includegraphics[width=\textwidth]{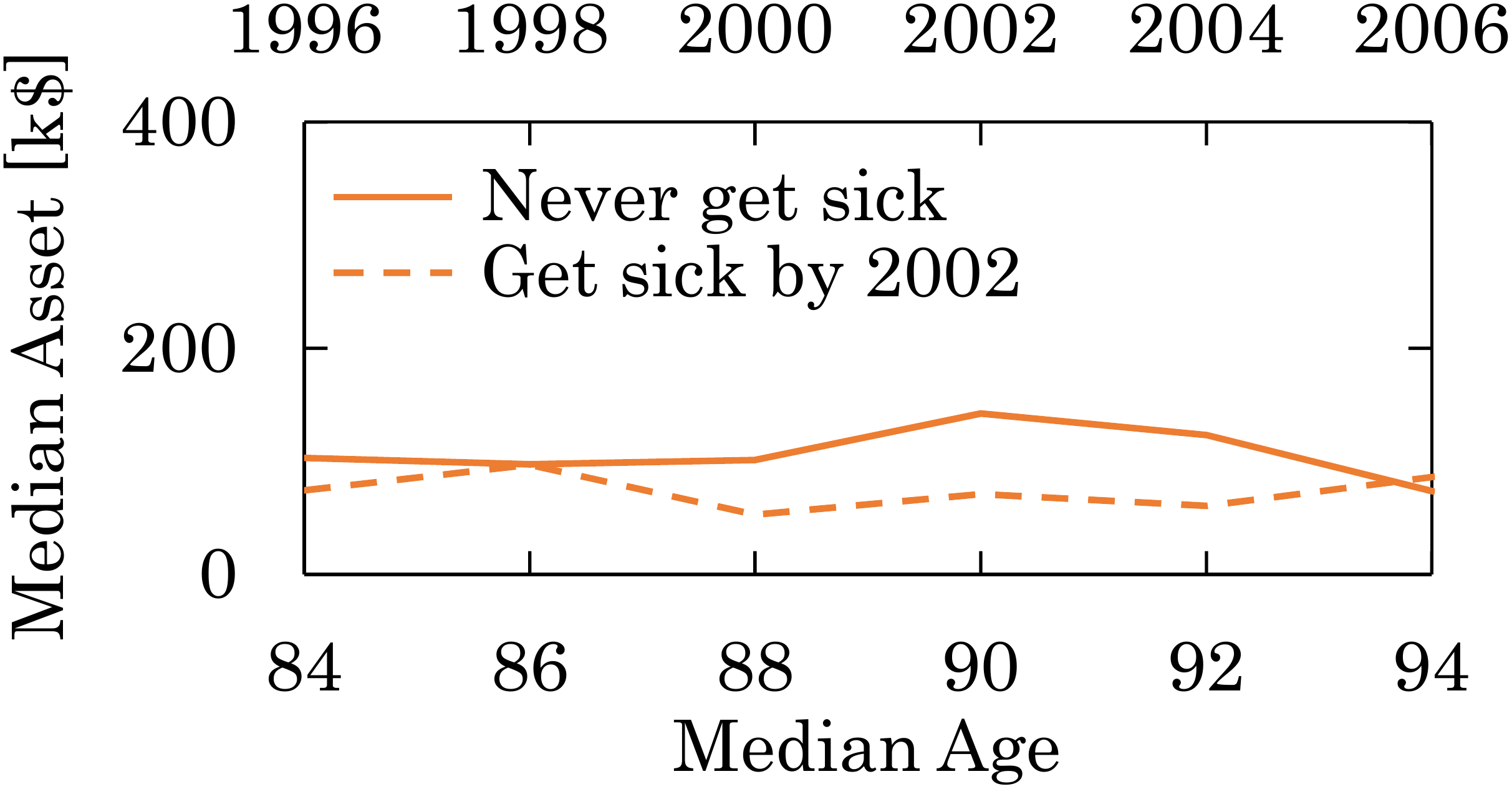}
\caption{Women's asset.} \label{fig:1b}
\end{subfigure}
\\\vspace{6pt}
\begin{subfigure}[t]{0.45\textwidth}
\centering \includegraphics[width=\textwidth]{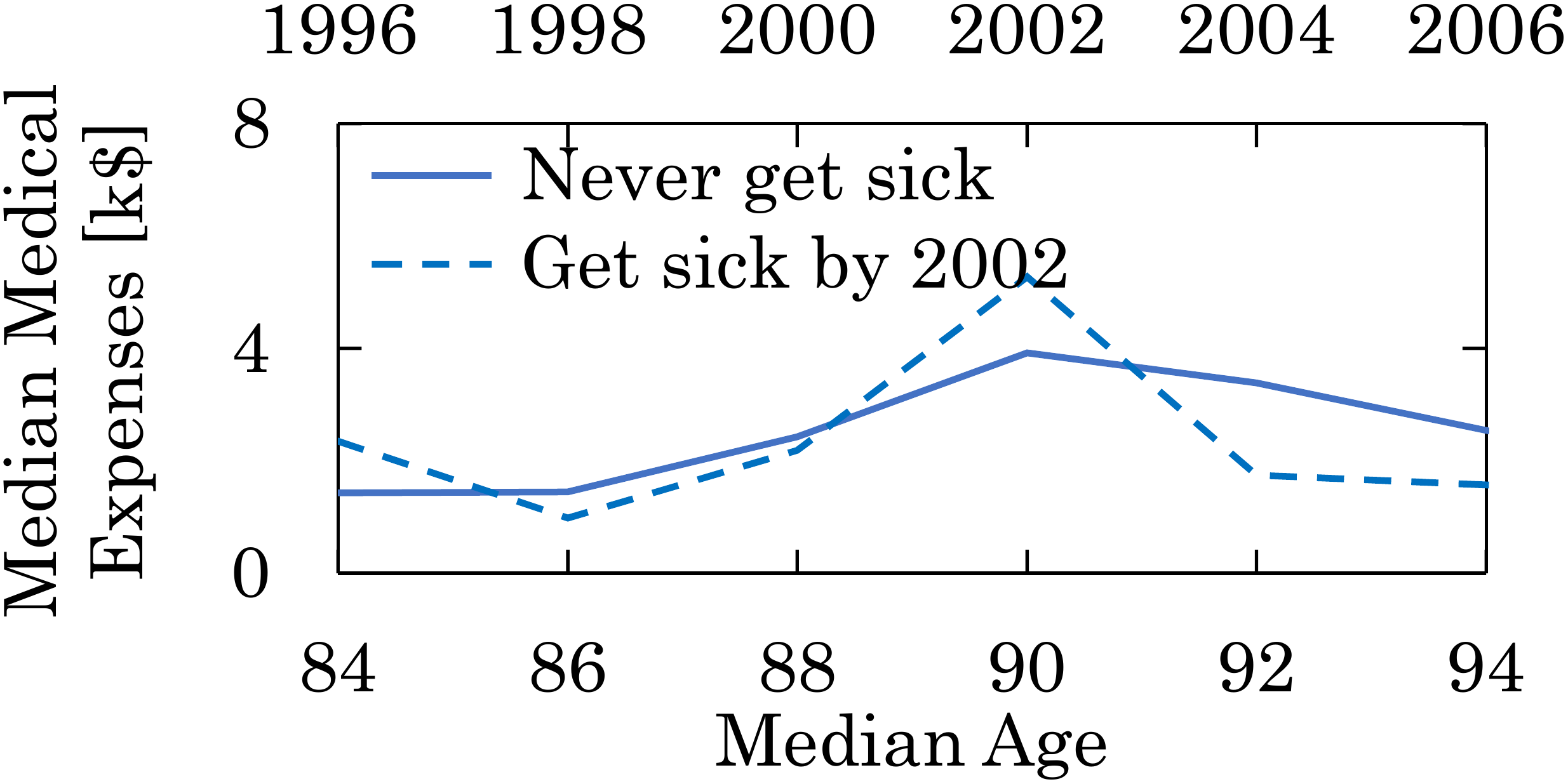}
\caption{Men's medical expenses.} \label{fig:1e}
\end{subfigure}
\qquad
\begin{subfigure}[t]{0.45\textwidth}
\centering \includegraphics[width=\textwidth]{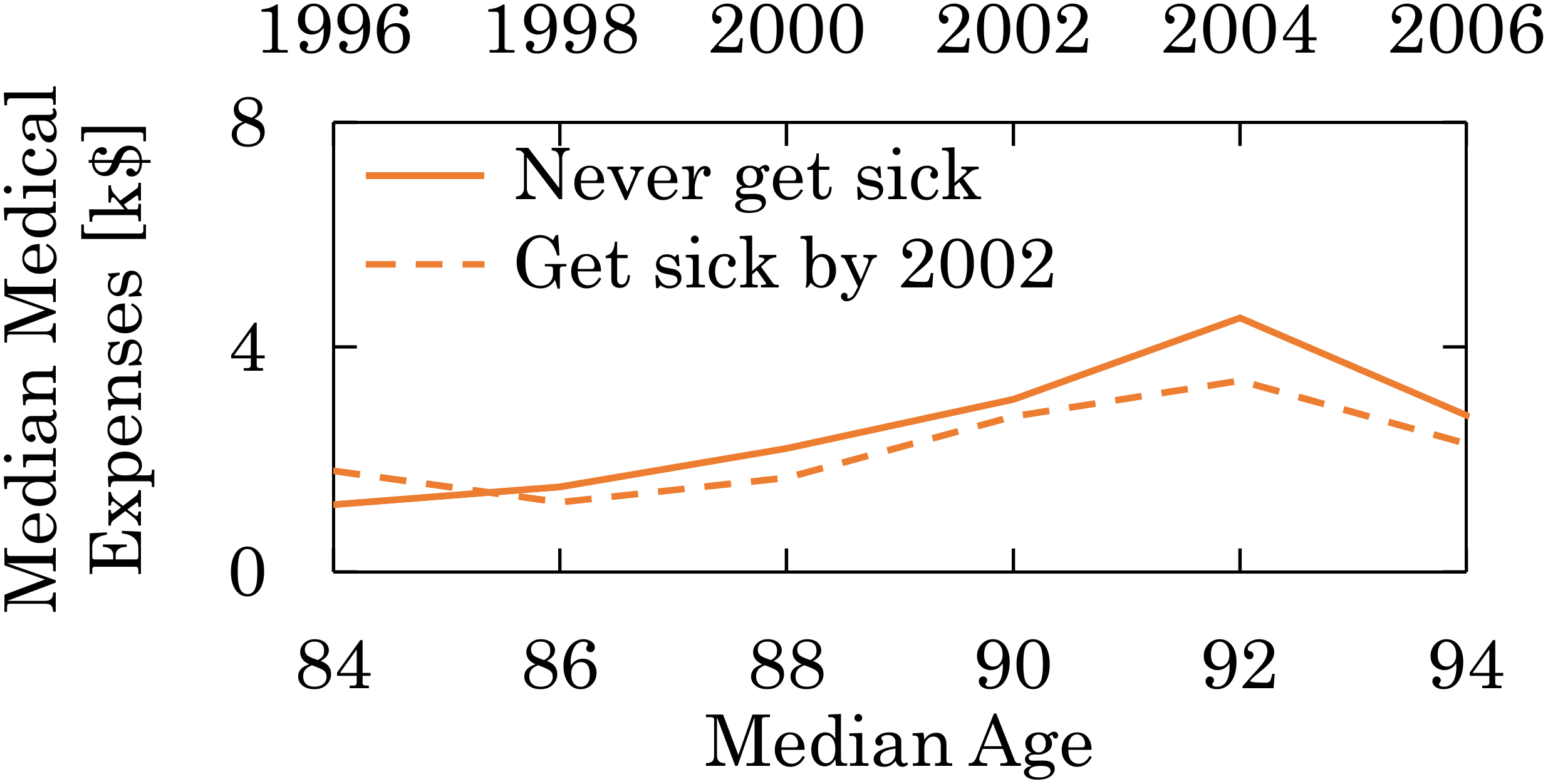}
\caption{Women's medical expenses.} \label{fig:1f}
\end{subfigure}
\\\vspace{6pt}
\begin{subfigure}[t]{0.45\textwidth}
\centering \includegraphics[width=\textwidth]{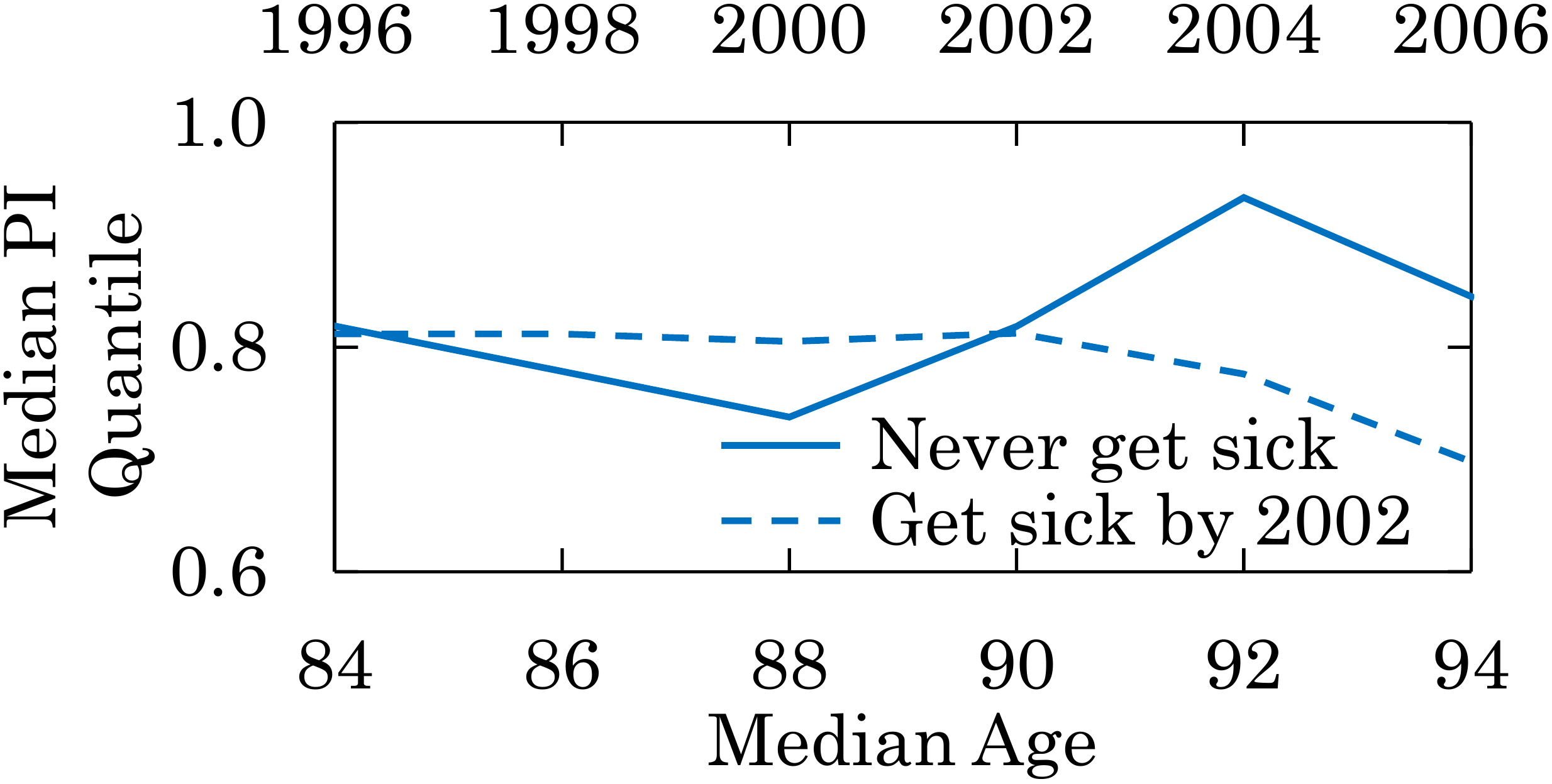}
\caption{Men's permanent income.} \label{fig:1c}
\end{subfigure}
\qquad
\begin{subfigure}[t]{0.45\textwidth}
\centering \includegraphics[width=\textwidth]{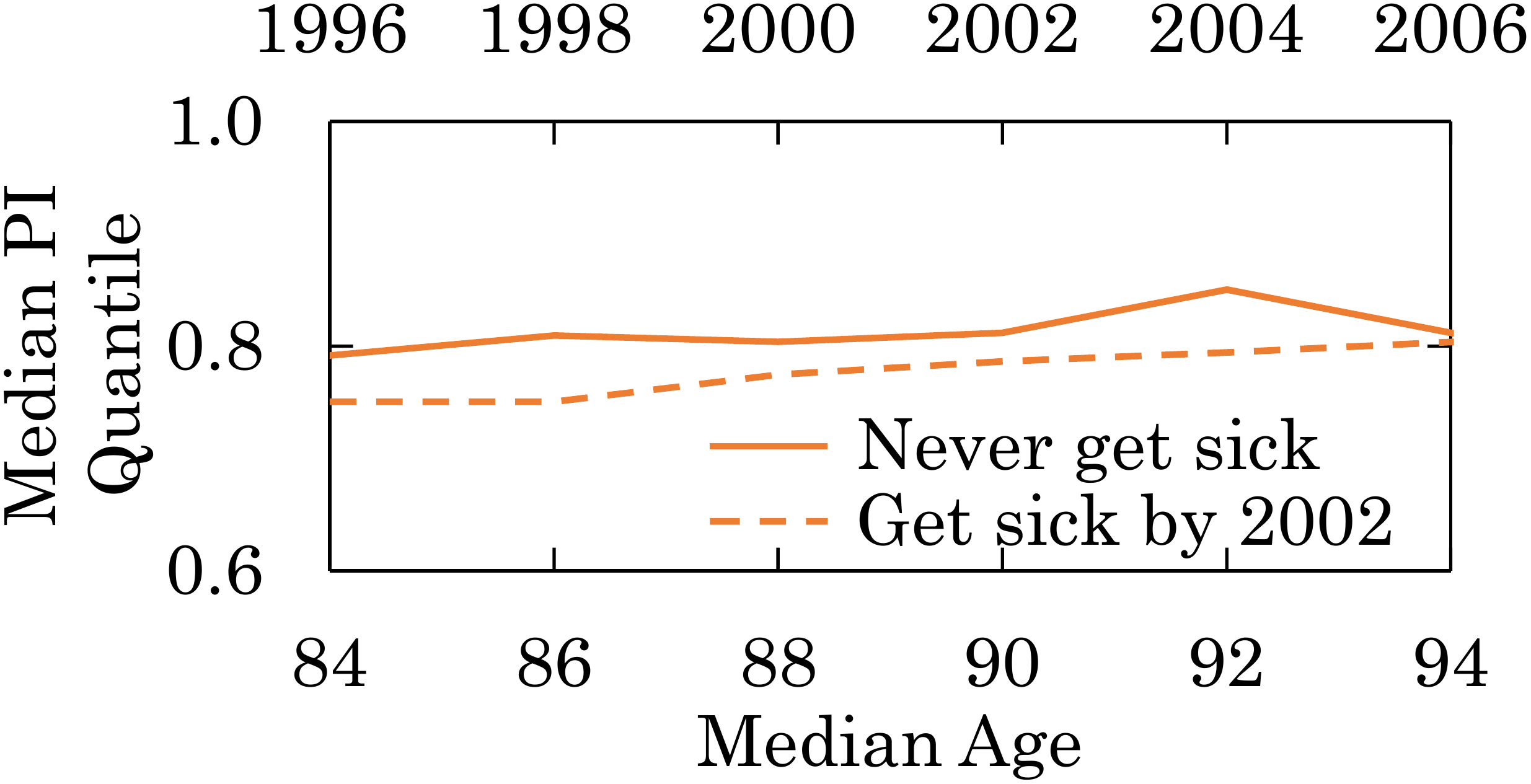}
\caption{Women's permanent income.} \label{fig:1d}
\end{subfigure}
\begin{center}
     \begin{minipage}{\textwidth} 
        \small \textit{Note:} \Cref{fig:1a,fig:1b,fig:1c,fig:1d,fig:1e,fig:1f} are for 4\textendash 5th PIqs in Cohort 3. Solid lines are for those who stay healthy for the duration of their observation; dashed lines for those who are healthy in 1996 and become unhealthy by 2002.
    \end{minipage}
\end{center}

\label{fig_health_asset}
\end{figure}

\Cref{fig:1g,fig:1h} are the proportions of individuals who survive for the next five years at ages 85 and 90, conditional on gender and health.
We see that the health status, along with gender, is a strong predictor of life expectancy in years when the medical expenditure soars.

Heterogeneity in the survival materializes as a difference in the savings.
\Cref{fig:1a,fig:1b} give the trajectories of the median assets for the 4th and 5th PI quintiles in Cohort 3.
The solid lines are those who were healthy throughout the survey periods and the dashed lines are those who were healthy in 1996 but reported unhealthy in 1998, 2000, or 2002.
We see that men who were exposed to the health shock (hence the survival shock) dig into their savings much more than healthy men. With higher survival rates, women exhibit the trend to a much lesser degree.

Such difference in the asset profiles seems to be driven neither by the difference in medical expenses nor by survival selection among the rich.
\Cref{fig:1e,fig:1f} show the median medical expenses during the same periods; we observe similar trajectories across gender and health.
\Cref{fig:1c,fig:1d} show the median PI quantiles of the survivors; if there is attrition of rich or poor individuals that affects the median assets, we expect to see a change in the median PI quantiles. However, they do not differ much by at least age 90 while bifurcation of the asset profiles begins at age 90.

These findings are suggestive that the difference in the asset profiles is attributable to the change in the saving behaviors. 
The health status changes the exposure to the medical expenditure risk through the survival probability, which then induces changes in the saving behavior by shifting the balance between the bequest motive and medical expenditure risk.

\section{Estimation}\label{sec:est}
We estimate this model by using the Adversarial Estimation methods introduced in \cite{kmp2023adversarial}. In the next subsection we introduce the estimation framework. We then discuss why this approach has merit in the context of estimation of saving motives for the elderly.

\subsection{Adversarial Estimation Framework}
The adversarial estimation has two main components: simulation and discrimination.
The simulation component is the same as SMM or indirect inference, but the discriminator component is new. 
The essence of the adversarial
framework is to find a parameter value for which the corresponding simulated
data is indistinguishable from the real data according to the discriminator. We
now describe each component in turn.

\subsubsection{Simulation}
Let $\{X_i\}_{i=1}^n$ denote i.i.d. draws from an unknown distribution $P_0$, the real data. In our context that would be the history of saving patterns for an individual, in addition to other characteristics, such as their PI quintile, their age or their gender, stack in a vector. This represents the data the structural model should match. 
Now, given a parameter $\theta$ governing the utility of agents, we want to construct a sample of simulated data. More specifically, we consider $\theta=(\nu, MPC, K)$, where $\nu$ is the risk aversion, $MPC$ is the marginal propensity to consume at the moment of death, and $K$ is the level above which agents derive utility of bequeathing. Then, for a given sample size $m$, we obtain a sample of simulated observations, $\{X_{i,\theta}\}_{i=1}^m$ after (a) numerically solving the model in Section \ref{sec:2}, and (b) generating a set of shocks that represent the randomness faced by the simulated agents, such as health shocks, income shocks, etc. We denote $P_\theta$ the distribution associated to the simulated sample.

\subsubsection{Discriminator}
The discriminator is the novelty in the estimation framework and is the key component in the construction of the objective function for the adversarial estimator.
For some $\theta$ and $x$, consider the problem of assessing whether $x$ is drawn from $P_\theta$ or $P_0$. That is, if $x$ ``looks like'' decisions that could have been taken by a real economic agent, or if instead looks like decisions that would could have been taken by an agent that takes decisions according to model $\theta$. 
If $P_\theta$ is very different from $P_0$, it should be easy to distinguish realizations of $P_\theta$ from those of $P_0$. If they are close, it should be harder.
The idea, therefore, is to pick a classification algorithm that takes a value $x$ and predicts which distribution it came from, and to search for the value of $\theta$ for which the algorithm can distinguish among the two distributions the least.\footnote{In the Indirect Inference framework the choice of discriminator can be thought of as the choice of auxiliary model (e.g. \cite{gourieroux1993indirect}.}

If we had access to the probability density functions corresponding to $P_0$ and $P_\theta$, it would be easy to assign the provenance of $x$ according to the likelihood of $x$ for each distribution. But those distributions are typically unknown in models involving dynamic optimization, since the value functions cannot be obtained in closed form. Instead, 
 we can take
advantage of the availability of samples $\{X_i\}_{i=1}^n$ and $\{X_{i,\theta}\}_{i=1}^m$ to estimate the
extent to which, for a given $\theta$, these two distributions are different. In particular,
we use the fitted predictions of a discrete choice model, the \textit{discriminator}, where the dependent variable is $1$ if the data is real and $0$ if it is
simulated, and the explanatory variables are $X_i$ if the data is real,
and $X_{i,\theta}$ if it is simulated. When $\theta$ is a poor candidate to describe the observed
data, the predictions will be either close to $1$ or close to $0$. However,
as $\theta$ becomes a better candidate to describe the real data, the distribution of
the prediction will concentrate around $0.5$.


The {\em adversarial estimator} is defined by the following minimax problem:%
\begin{equation} \label{eq:main_estim}
\hat{\theta}=\argmin_{\theta\in\Theta}\max_{\vphantom{\theta}D\in\mathcal{D}_n} \frac{1}{n}\sum_{i=1}^n\log D(X_i)+\frac{1}{m}\sum_{i=1}^m\log(1-D(X_{i,\theta})),
\end{equation}
where $\mathcal{D}_n$ is the set of possible discriminator. Examples of $\mathcal{D}_n$ could be standard discrete choice models such as logit, or probit, but also more flexible models, such as regression trees or neural networks. The more flexible the discriminator, the more features of the data will be used to compare the two datasets. 
%
%

Since $D$ is between $0$ and $1$, both $\log D$ and $\log(1-D)$ are nonpositive.
If $\{X_i\}$ and $\{X_{i,\theta}\}$ are very different from each other, the discriminator may be able to find $D$ that assigns $1$ on the support of $\{X_i\}$ and $0$ on the support of $\{X_{i,\theta}\}$, in which case the inner maximization attains the value of zero.
In general, therefore, the inner maximization will give a number between $2\log(1/2)$ and $0$, and the closer it is to $2\log(1/2) \approx -1.3863$, the less able the discriminator is to classify the observations.

The asymptotic distribution of the adversarial estimator depends on the choice of $\mathcal{D}_n$.
If the discriminator is logistic, the asymptotic variance of the adversarial estimator coincides with SMM (see Section S.1 in the Supplementary Appendix of \cite{kmp2023adversarial}).
If $\mathcal{D}_n$ is a nonparametric discriminator, and under suitable conditions, the asymptotic variance will be the same as the Maximum Likelihood Estimator, and hence, the adversarial method will deliver the most precise estimator (see Section 4.2, Corollary 4 in \cite{kmp2023adversarial}).


\subsection{Why Using the Adversarial Method?}
In this section we argue that using the Adversarial Method to learn about why do the elderly save is a good idea. Indeed, given that identification of the different mechanisms of savings is so challenging, having a method that (a) provides precise estimates as accurate as possible, (b) can automatically pick the features of the data that are most relevant for identification of the different mechanisms, is very attractive. 

As discussed in \cref{sec:1}, the difficulty of identification for savings mechanisms is challenging because bequests motives are more likely among the wealthy, which in turn have the most incentives to accumulate wealth in anticipation of high medical expenses as they expect to live into their 80's or 90's. In the context of SMM it is hard to specifically pick moments that are targeted at separating these two channels. The adversarial method avoids having to pick specific moments and instead lets the discriminator select the aspects of the data that are most informative.

As discussed in \cref{sec:id}, shocks in life expectancy, conditional on gender, could provide additional identifying variation to separate precautionary savings motives from bequests motives. A naive approach would be to expand the set of moments considered by DFJ by conditioning on health histories and sex. However, there are many possible health histories, and with a limited sample size, those conditional moments will be quite  poorly estimated, hence compromising any possible gain in efficiency.\footnote{See for example \cite{newey2004higher} for an analysis of bias arising from large number of moments in 2-step SMM/GMM.} Instead, in the adversarial method, including information on health amounts to enlarge the set of inputs that the discriminator will use to tell apart both samples.

Finally, a technical reason: the adversarial method, when combined with a neural network discriminator, can deliver even more precise estimators when $X_i$, the data that used the feed the discriminator, can be represented as a lower dimensional vector. That is, as opposed to SMM, where the number of moments grows exponentially with the values of the conditioning variables, the quality of the adversarial estimator only depends on the underlying dimensionality of the set of inputs. In the supplemental appendix, section \ref{sec:autoencoder} we provide a heuristic way to check low\hyp{}dimensionality using {\em autoencoders}. Low\hyp{}dimensionality is a feature of structural models and economic data, where a small number of factors drives variation across multiple outcomes \citep[e.g.,][]{cunha2010estimating}. In our data, we find that a 13-dimensional vector, containing individual health and survival histories, can be represented by only 4 dimension.

\subsection{Implementation} \label{sec:practical}

Following \citetalias{dfj}, we carry out estimation in two steps: (1) estimate $\pi_H$, $\pi_S$, $m$, $\sigma$, $\rho_m$, $\sigma_\xi$, $\sigma_\epsilon$ (in fact, we borrow numbers from \citetalias{dfj}), (2) estimate $\nu$, $\text{MPC}$, and $k$ using our adversarial approach. The parameters $r$, $\tau$, $\tilde{\tau}$, and $\tilde{x}$ are fixed as in the original paper, and $\beta = 0.971$.
For $\underline{c}$, we fix it at \$4,500 to reflect annual social security payments.\footnote{In their preferred specification \citetalias{dfj} estimate $\beta$ and $c_{floor}$,  in addition to $\nu$, $MPC$ and $k$. Instead, we fix $\beta$ and $c_{floor}$ to a reasonable value according to the literature. Sensitivity analysis shows that changing $\underline{c}$ mostly affects the risk aversion parameter.}
After the second step, we can also recover $\vartheta$ and $\underline{a}$.
In this section we discuss implementation details of the adversarial method. 

\subsubsection{Choice of Inputs and Discriminators} \label{sec:choice}

The method requires the choice of inputs $X_i$ and the choice of the discriminator $\mathcal{D}$. 

We consider two different sets of inputs to the discriminator.
The first set consists of the log age of an individual in 1996, permanent income (the aforementioned proxy), the profile (full history) of asset holdings, and the profile of survival indicators,%
\footnote{All individuals are alive in 1996, so we drop $s_{t_{1996}}$.}

\[
	X_1\vcentcolon=(1,\log t_{1996},I,a_{t_{1996}},\dots,a_{t_{2006}},s_{t_{1998}},\dots,s_{t_{2006}})\in\mathbb{R}^{14}.
\]
This is intended to capture similar identifying variation as \citetalias{dfj}.
The second set is augmented with gender and the profile of health status,
\[
	X_2\vcentcolon=(X_1,g,h_{t_{1996}},\dots,h_{t_{2006}})\in\mathbb{R}^{21},
\]
aiming to capture identifying variation to disentangle bequest motive from precautionary savings. 

The choice of the discriminator is more nuanced, but the goal is to choose a discriminator that strikes a balance between flexibility, to precisely distinguish between the simulated and the real data samples, and parsimony to avoid overfit. 

We use cross validation to choose the discriminator.
We focus on feed\hyp{}forward neural networks with sigmoid activation functions with at most two hidden layers.
We fix $\theta$ at a preliminary estimate; split the actual data into sample 1 (80\%) and sample 2 (20\%); estimate $D$ with sample 1, varying the numbers of nodes and layers; evaluate their classification accuracy with sample 2;%
\footnote{We use the classification accuracy provided by Keras's ADAM, which is based on thresholding.}
pick the network configuration with the highest accuracy.
The selected neural network discriminator consists of two hidden layers, the first with 20 nodes and the second 10 nodes.




\subsubsection{Algorithm}\label{sec:computation}

We consider an iterative algorithm that solves the optimization problem in (\ref{eq:main_estim}).

\begin{alg}[Estimation]\label{alg:estim} \hfill
\begin{enumerate}[i.]
	\item Initialize $\theta=\theta^{(0)}$. Fix a set of random shocks $\{\tilde{X}_i\}_{i=1}^m$ and any random seed if stochastic optimization is used.
	\item For given $\theta = \theta^{(s)}$, generate $\{X_i^{\theta^{(s)}}\}_{i=1}^m$ using $\{\tilde{X}_i\}_{i=1}^m$. \label{alg:est:2}
	\item Train the discriminator with $\{X_i\}_{i=1}^n$ and $\{X_i^{\theta^{(s)}}\}_{i=1}^m$.
	\item Compute the gradient $\Delta(\theta^{(s)})$ of the objective function with respect to $\theta$.
	\item Set $\theta^{(s+1)} = \theta^{(s)} - \xi \Delta(\theta^{(s)})$ where $\xi>0$ is a learning rate. \label{alg:est:5}
	\item Repeat (\ref{alg:est:2})\textendash (\ref{alg:est:5}) until $\Delta(\theta)\approx0$.
\end{enumerate}
\end{alg}

To train the neural network discriminator, we make use of off\hyp{}the\hyp{}shelf routines in the R Keras package.
They come with implementations of various techniques such as back\hyp{}propagation, automated differentiation, and stochastic gradient descent. See \cref{sec:estimation_al} in the Supplementary Appendix for full details on the algorithm implementation and computational specifications.

The algorithm may get stuck in a local minimum.
It is advised to use several different initial values to explore a wide space.






\subsubsection{Inference}

The asymptotic variance formula of the adversarial estimator is challenging to estimate since we do not have the closed\hyp{}form likelihood.
We propose to use the bootstrap to estimate confidence intervals either based on bootstrap variance or based on bootstrap quantiles.\footnote{When standard bootstrap is computationally burdensome, we suggest using a fast version of the bootstrap, as proposed by \citet{honore2017poor}.} 

\begin{alg}[Bootstrap] \hfill
\begin{enumerate}[i.]
	\item Let $\{X_i^\ast\}_{i=1}^n$ and $\{\tilde{X}_i^{\theta\ast}\}_{i=1}^m$ be the bootstrap samples of actual and synthetic observations of sizes $n$ and $m$, drawn randomly with replacement. \label{alg:boot:1}
	\item Solve (\ref{eq:main_estim}) with $\{X_i^\ast\}_{i=1}^n$ and $\{\tilde{X}_i^{\theta\ast}\}_{i=1}^m$ to obtain a bootstrap estimator $\hat{\theta}_{n,m}^{\ast(1)}$. \label{alg:boot:2}
	\item Repeat (\ref{alg:boot:1})\textendash (\ref{alg:boot:2}) for $S$ times to obtain $S$ bootstrap estimators $\{\hat{\theta}_{n,m}^{\ast(1)},\dots,\hat{\theta}_{n,m}^{\ast(S)}\}$.
	\item Use the distribution of $\{\hat{\theta}_{n,m}^{\ast(s)}\}_{s=1}^S$ to approximate the distribution of $\hat{\theta}_{n,m}$.
\end{enumerate}
\end{alg}


\section{Monte Carlo}\label{sec:MC}
In this section we report results of a Monte Carlo experiment to assess the performance of the adversarial method in this model. We also use this exercise to show how the iterative algorithm outlined in \cref{sec:computation} successfully manages to solve the minimization problem (\ref{eq:main_estim}). 

The set-up is as follows: we simulate 20,000 individuals, in an economy where they take savings decisions according to the model in \cref{sec:2}. Preference parameters are fixed at $\nu = 3.8$, $MPC = 0.25$, and $K_0 = 10,000$.\footnote{These parameter values have been chosen arbitrarily.} These correspond to a moderate bequest motive, with an MPC at death of 0.25 cents per dollar, and an  implied asset floor of $\$3,266$. We generate 100 of these economies and take them as the ``real'' data. 

For each generated sample we run 4 different estimators defined by two different choices of inputs and two different types of discriminators. More specifically, we use either $X_1$ or $X_2$ as inputs, and either 1-layer NN with 5 nodes and 1-layer NN with 10 nodes as discriminators (herein we call them simple discriminator and sophisticated discriminator, respectively). With these 4 specifications we are looking to study 2 main things. First, comparing specifications that use $X_1$ or $X_2$ allows us to test the hypothesis that health and gender information provides useful variability for precise estimation of bequest motives. Recall that input $X_1$ contains information only on savings patterns, PI quantiles, and cohort, while $X_2$ adds information on health status as well as gender. Second, we are looking to compare the precision of estimates across discriminators that vary in complexity. A 10 node neural network discriminator has the potential to capture more features of the data than a 5 node discriminator and lead to more precise estimates. However, it also represent overall more parameters to estimate, and they could also end up diminishing precision.


\subsection{Algorithm Performance}
The algorithm alternates between training the discriminator, and updating the parameter values in the direction of numerically estimated gradients. The training of the discriminator is done using stochastic gradient methods, which can make the objective function wiggly due to additional randomness in the training process. In order to make the objective function as smooth as possible, we train the discriminator thoroughly. While this slows down computation, and it might lead to overfit occasionally, it provides the most reliable way to obtain parameter estimates, in our experience. 

\begin{figure}[htbp]
 \caption{Monte Carlo Performance}
    \centering
    \includegraphics[width=\textwidth]{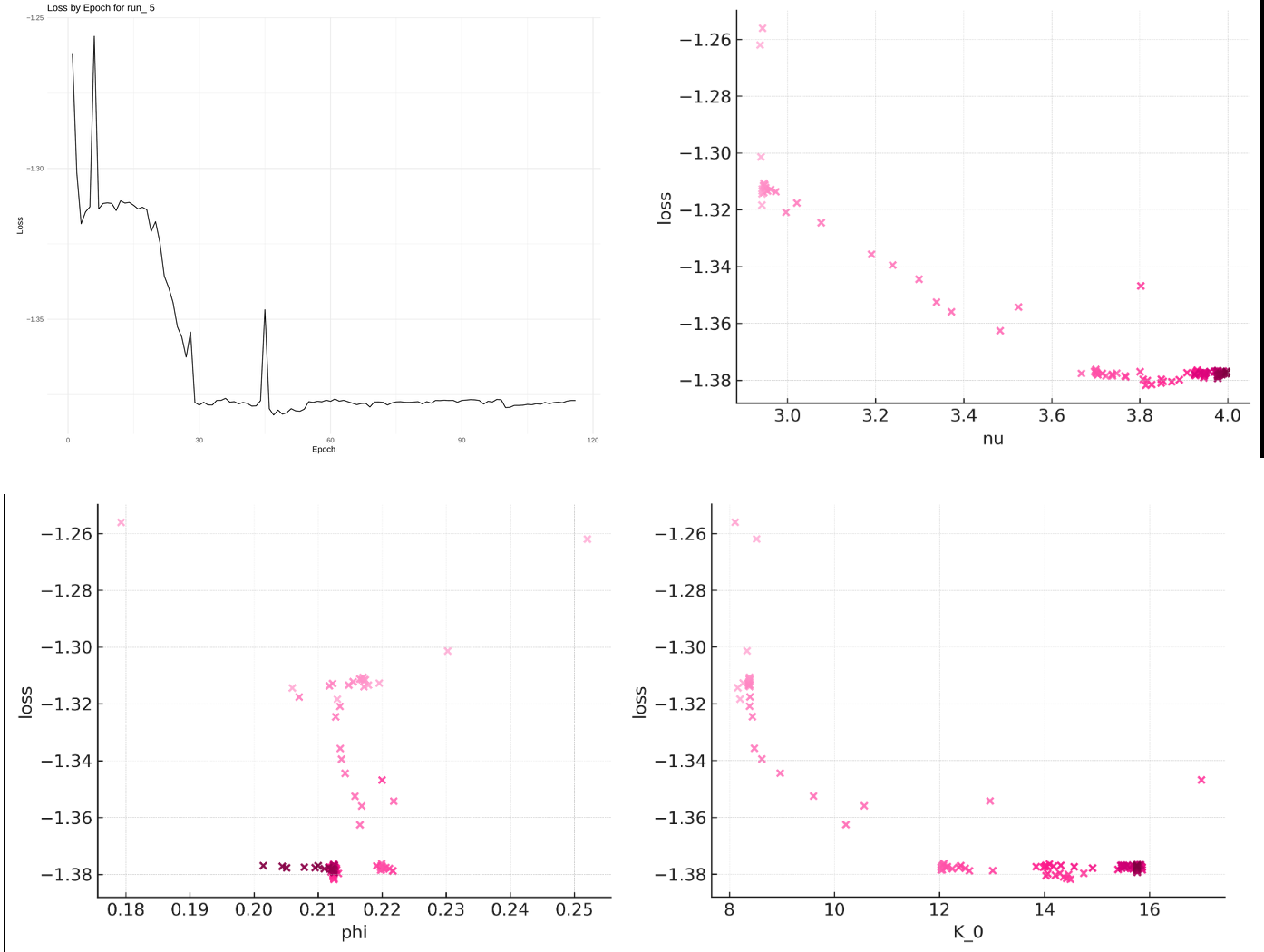}
    \label{fig:loss_vs_params_2x2}
  \begin{minipage}{\textwidth} 
        \small \textit{Note:}  Top-left: loss by iteration for one run. 
        Top-right: trajectory for $\nu$ across iterations vs loss value. 
        Bottom-left: trajectory for $\phi$ across iterations vs loss value. 
        Bottom-right: trajectory for $K_0$ across iterations vs loss.
        Colors indicate iteration number (from light to dark).
    \end{minipage}

\end{figure}

In the upper left panel of Figure \ref{fig:loss_vs_params_2x2}, we show the evolution of the loss function for a particular experiment, with $X_2$ as inputs, and the sophisticated discriminator. Across iterations, the loss function decreases steadily, except for a couple of instances where the training of the NN seems to overfit.\footnote{Dropout techniques can be added to the training of the discriminator to mitigate it, although it did not seem to make a difference in the performance of the estimator in finite samples.} Nonetheless, the loss function eventually reaches values close to -1.38, the theoretical minimum, before the gradient is sufficiently close to 0 (condition for stopping) and the algorithm stops. In addition, we show the evolution of the parameter estimates across iterations in the three remaining panels. The iterations are color coded, where parameter values of the earlier iterations are lighter than the ones obtained at the last iterations. The graphs show that as iterations evolve the parameter values approach the true parameter values, and for the lowest values of the loss they set into values close to the true ones. 

\subsection{Recovering the Parameters}

Table \ref{tab:mc_nn_results_x1x2} collects results on the 4 designs. Panel A shows results when using $X_1$ as input, while Panel B shows results when using $X_2$. Within each panel, the three left columns show results using the simple discriminator, and the three right columns show results using the sophisticated discriminator.

We start by drawing attention to the mean value of the estimates: Across all 4 designs the mean estimates match closely the "true parameters", confirming that, under correct specification, the estimator is unbiased.

We now turn to assess variability of estimates across MC simulations. Let us first focus on Panel A, the estimates corresponding to the simple discriminator and $X_1$ as input. Estimates are recovered with significant precision for all parameters, especially for $\nu$ and $\phi$, and to a lesser extent, $K_0$. These results mirrors the informal observation that in Figure \ref{fig:loss_vs_params_2x2} the range of values that are close to the minimum loss is the largest for $K_0$, and less so for $\nu$ and $MPC$. In addition to the variance we report the quantile bootstrapped confidence intervals for $5\%$ and $10\%$ confidence level. True parameter values coincide at the midpoint of the interval, suggesting that the distribution is symmetric around the truth. The confidence intervals based on the normal approximation would had been wider, suggesting that the variance might be inflated due to computational reasons.

We now turn to compare the precision of the estimates for $X_1$ when using a more sophisticated discriminator. Interestingly, the variance for $\nu$ and $K_0$ increase relative to the simple discriminator. The variance of $MPC$ remains practically unchanged. The comparison in terms of interquantile ranges is mostly consistent with the behavior of the variance. Overall, there is not a clear improvement to use a more sophisticated discriminator when using $X_1$. 

In terms of results for $X_2$, we start by comparing the precision of the estimator across discriminators: in contrast with the results with $X_1$, the variance of the estimators significantly decrease when using the sophisticated discriminator. This is consistent with the main theoretical result in \cite{kmp2023adversarial}, where we show that as the number of inputs increases, the complexity of the discriminator also needs to increase in order to capitalize on the gains of adding information to the estimation. In fact, the adversarial estimator can eventually achieve the same precision as the Maximum Likelihood Estimator, the most efficient estimator.\footnote{It is difficult to know ex-ante if the discriminator we are using is sufficiently flexible. We recommend trying with multiple discriminators to assess the robustness and the precision of the estimates.} However, when the discriminator is not sufficiently flexible, the estimator is still consistent, but its variance will be larger.\footnote{For a more formal discussion on the effect of using a not-so-flexible discriminator, see Section 4.4. in \cite{kmp2023adversarial}.} The interquantile ranges also decrease across the board when using the more sophisticated discriminator.

Now, in terms of the comparison between an estimator that uses $X_1$ vs $X_2$: the estimator that uses $X_2$ with the sophisticated discriminator is overall more precise than the one that uses $X_1$. More specifically, the variances and interquantile ranges are smaller for $\nu$ and $MPC$, although they are slightly larger for $K_0$, when using $X_2$. The precision of the asset floor improves when using $X_2$ as opposed to $X_1$, but it is a bit less clear for $\theta$. This provides strong evidence that the use of gender and health shocks are an important source of identifying variation for the bequests motives savings mechanism.

We finish by reporting on the shape of the loss function. Figure \ref{fig:quad_form} in the Appendix provides evidence of the quadratic shape of the loss around the true parameter values. This provides evidence that the estimator is likely to be well approximated by a Gaussian distribution.

\begin{table}[htbp]
\centering
\caption{Monte Carlo Results}
\label{tab:mc_nn_results_x1x2}

\textbf{Panel A: Input $X_1$}\\[3pt]
\begin{tabular}{lcccccc}
\toprule
 & \multicolumn{3}{c}{\textbf{1L NN (5 nodes)}} & \multicolumn{3}{c}{\textbf{1L NN (10 nodes)}} \\ 
\cmidrule(lr){2-4} \cmidrule(lr){5-7}
 & $\nu$ & $MPC$ & $K_0$ & $\nu$ & $MPC$ & $K_0$ \\
\midrule
Mean & 3.829 & 0.249 & 10.42 & 3.840 & 0.252 & 10.37 \\
Variance & 0.033 & 0.0008 & 4.28 & 0.054 & 0.0007 & 4.40 \\
\midrule
quantile & \multicolumn{6}{c}{} \\
\quad 2.5\%  & 3.356 & 0.192 & 6.25 
& 3.247 & 0.205 & 6.33 \\
\quad 97.5\% & 4.194 & 0.307 & 14.97 
& 4.585 & 0.311 & 15.00 \\
\quad Range  & 0.838 & 0.115 & 8.72 
& 1.338 & 0.106 & 8.68 \\
\midrule
quantile & \multicolumn{6}{c}{} \\
\quad 5\%  & 3.476 & 0.205 & 7.29 
& 3.482 & 0.215 & 7.05 \\
\quad 95\% & 4.141 & 0.298 & 14.20 
& 4.206 & 0.300 & 14.33 \\
\quad Range & 0.665 & 0.093 & 6.91 
& 0.724 & 0.085 & 7.28 \\
\bottomrule
\end{tabular}
\vspace{1.2em}

\textbf{Panel B: Input $X_2$}\\[3pt]
\begin{tabular}{lcccccc}
\toprule
 & \multicolumn{3}{c}{\textbf{1L NN (5 nodes)}} & \multicolumn{3}{c}{\textbf{1L NN (10 nodes)}} \\ 
\cmidrule(lr){2-4} \cmidrule(lr){5-7}
 & $\nu$ & $MPC$ & $K_0$ & $\nu$ & $MPC$ & $K_0$ \\
\midrule
Mean & 3.848 & 0.258 & 10.66 & 3.820 & 0.247 & 10.80 \\
Variance & 0.0516 & 0.0047 & 7.538 & 0.0132 & 0.0006 & 6.168 \\
\midrule
quantile & \multicolumn{6}{c}{} \\
\quad 2.5\%  & 3.345 & 0.183 & 7.34 & 3.605 & 0.198 & 6.94 \\
\quad 97.5\% & 4.311 & 0.315 & 18.39 & 4.063 & 0.297 & 16.88 \\
\quad Range  & 0.966 & 0.132 & 11.05 & 0.458 & 0.099 & 9.94 \\
\midrule
quantile & \multicolumn{6}{c}{} \\
\quad 5\%  & 3.419 & 0.199 & 7.49 & 3.645 & 0.207 & 7.47 \\
\quad 95\% & 4.215 & 0.299 & 16.12 & 4.001 & 0.289 & 15.00 \\
\quad Range & 0.796 & 0.100 & 8.63 & 0.355 & 0.082 & 7.52 \\
\bottomrule
\end{tabular}
\vspace{0.75em} 

\begin{minipage}{0.95\textwidth}

        \small \textit{Note:} Panel A shows results when using $X_1$ as input, while Panel B shows results when using $X_2$.1L NN (5 nodes) is the simple discriminator, 
        and 1L NN (10 nodes) is the sophisticated discriminator. For each MC sample we initialize the algorithm with a single initial condition randomly drawn from a normal centered at the true parameter, with variances 1, 0.025, and 4, respectively for $\nu$, $MPC$, and $K_0$. 
    \end{minipage}
\end{table}

\subsection{Bootstrap Performance}
We now provide evidence that the Bootstrap is a reliable way to characterize the distribution of the estimator. Table \ref{tab:bootstrap_results} collects results on 50 bootstrap replications of a particular MC run with $X_2$ as input and simple discriminator.\footnote{We pick arbitrarily one particular design for assessing the bootstrap.} We initialize each bootstrap replication estimation with the same random initialization as in the MC experiments. We are going to compare the distribution of the bootstrap to the distribution obtained with the MC experiments. 

The mean across bootstrap replications match well the true parameters. In terms of variance, the bootstrap variance of $\nu$ and $MPC$ is in the ballpark of the MC one, while it underestimates the one of $K_0$. In terms of interquantile range, the bootstrap closely matches the interquantiel ranges for $\nu$ and $\phi$, while it underestimates the one of $K_0$.

Overall, we conclude that the bootstrap provides a reliable method to characterize the variability of the estimator. 

\begin{table}[htbp]
\centering
\caption{Bootstrap Results for Parameter Estimates}
\label{tab:bootstrap_results}
\begin{tabular}{lccc}
\toprule
 & $\nu$ & $MPC$ & $K_0$ \\
\midrule
\textbf{True parameters} & 3.800 & 0.250 & 10.000 \\
\midrule
Mean & 3.831 & 0.257 & 10.37 \\
Variance & 0.0724 & 0.0020 & 6.58 \\
\midrule
quantile & \multicolumn{3}{c}{} \\
\quad 2.5\%  & 3.455 & 0.195 & 6.219 \\
\quad 97.5\% & 4.449 & 0.362 & 15.866 \\
\quad Range  & 0.994 & 0.167 & 9.647 \\
\midrule
quantile & \multicolumn{3}{c}{} \\
\quad 5\%  & 3.504 & 0.202 & 6.756 \\
\quad 95\% & 4.288 & 0.310 & 13.219 \\
\quad Range & 0.784 & 0.108 & 6.463 \\
\bottomrule
\end{tabular}

\vspace{0.75em} 

\begin{minipage}{0.95\textwidth}
        \small \textit{Note:} Results for 50 bootstrap replications using $X_2$ as inputs and the simple discriminator. We take out 3 replications that did not converge. 
    \end{minipage}
\end{table}

\section{Results}\label{sec:res}

In this section we put the adversarial method to work with the sample of elderly individuals originally used by \citetalias{dfj}. We first present results using the adversarial method, and we then compare them to the results obtained in \citetalias{dfj}. We conclude by showing the fit of the model and some counterfactual analysis to evaluate its economic implications.

\subsection{Results for the Adversarial Estimates}

Table \ref{tab:results_ad} collects parameter estimates of the adversarial method as well as bootstrap quantile confidence intervals at the $5\%$ level. We consider two specifications: one with inputs $X_1$, where the varaibility used is similar to that of \citetalias{dfj}, and $X_2$, where we also add information on gender and health trajectories, to add identifying variation for bequest motives in our estimates. We will use the Bootstrap to obtain confidence intervals. 

For both specifications we use a sophisticated discriminator, with a 2-layer neural network with dropout. Also for both estimations we fix the discount rate at 0.97, and the gross rate of return at 1.02. We also keep fixed the consumption floor at $\$4,500$ dollars, a number considered reasonable in the literature.

We start by looking at results for $X_1$. The risk aversion parameter, which governs both the utility of own consumption as well as the utility of bequeathing, is in the higher end, but still within range, of the literature. This parameter is well estimated, as can be seen from the narrow interquantile range, and is consistent with an environment where there is a high level of insurance, given the high consumption floor, and yet the decrease in savings is slow. We next turn to assess the MPC at death parameter, which can also be seen as the counterpart of the marginal propensity of bequeathing. A value of 0.2 means that, at the time of death, the elderly are willing to give $80$ cents of a dollar to their heirs, which could be considered substantial. However, bequests only operate above a threshold, which is jointly determined by $k$, $\nu$, and the MPC, the asset floor, $\underline{a}$. We estimate it at $\$4,243$, which represent the 24th percentile of the empirical distribution of assets one period before death in the sample. The magnitude of the parameters $k$ and $\theta$, the intensity and the curvature of the bequest motive, are harder to interpret, so we prefer to focus on interpreting the asset floor, and the MPC, which are deterministic transformations of the former ones. Nonetheless, the large variability in $\theta$ is mechanical as $\theta \propto 1/MPC^{\nu}$. Hence, values of MPC closer to 0 bring $\theta$ to vary widely. All together, these estimates convey that bequests motives are non-negligible, as they imply that a significant proportion of individuals across the wealth distribution obtain utility from bequeathing.

We now turn to the results for $X_2$: as in $X_1$, the risk aversion parameter is also relatively high while equally well estimated. In terms of MPC, we see a decrease in the marginal propensity of consumption at death from 0.20 to 0.12 together with a tightening of the inter-quantile range (from 0.215 to 0.171). This means that out of each dollar, agents are willing to bequeath 0.88 cents. In addition, we also see a decrease of the asset floor estimate, where now it is $\$1,320$. This represents the 22nd percentile of the distribution of assets one period before death. This decrease is partly driven by a substantial decrease in the estimate of the curvature of the bequest motive, $k$. The precision in estimation of the asset floor also increases substantially relative to estimates in $X_1$ (with the interquantile range going down from $5,347$ to $3,711$).

Overall, the estimates using $X_2$ strengthen the message that bequests are an important motive for saving for a large proportion of the elderly, and not only the very wealthy ones. The addition of information on health trajectories and gender not only imply a larger bequest motive, but also increase the precision of the estimates of MPC and asset floor. 

Finally, the last column in \ref{tab:results_ad} reports the cross\hyp{}entropy loss of each set of parameter estimates.\footnote{
To make a fair comparison, we take each set of estimates of $X_1$ and $X_2$ and solve the inner maximization of (\ref{eq:main_estim}) using $X_2$ as the input.}
The respective losses are -0.67 and -0.87 for inputs $X_1$ and $X_2$. This indicates that the parameters obtained in $X_2$ fit the data better, which makes $X_2$ our preferred specification.

\begin{table}[htbp]
\caption{Estimates for the Adversarial Method}
\label{tab:results_ad}
\centering
\small
\setlength{\tabcolsep}{5pt} 
\begin{tabular}{lcccccc}
\toprule
 & $\nu$ & $\vartheta$ & $k$ [k\$] & MPC & $\underline{a}$ [\$] & Loss \\
\midrule
\textbf{Adversarial $X_1$} & & & & & & \\
\quad Estimate & 6.14 & 4,865 & 16.89 & 0.20 & 4,243 & $-0.67$ \\
\quad 2.5\% Quantile & 5.12 & 191.81 & 14.10 & 0.087 & 1,941 & \\
\quad 97.5\% Quantile & 6.96 & 2.29$\times$10$^{6}$ & 21.96 & 0.302 & 7,288 & \\
\quad Interquartile range & 1.83 & 2.29$\times$10$^{6}$ & 7.85 & 0.215 & 5,347 & \\
[2pt]
\textbf{Adversarial $X_2$} & & & & & & \\
\quad Estimate & 5.99 & 1.93$\times$10$^{5}$ & 10.02 & 0.12 & 1,320 & $-0.78$ \\
\quad 2.5\% Quantile & 4.95 & 763.69 & 7.36 & 0.077 & 804.14 & \\
\quad 97.5\% Quantile & 6.81 & 7.66$\times$10$^{6}$ & 21.88 & 0.248 & 4,576 & \\
\quad Interquartile range & 1.86 & 7.66$\times$10$^{6}$ & 14.51 & 0.171 & 3,711 & \\
\bottomrule
\end{tabular}
\begin{center}
     \begin{minipage}{\textwidth} 
        \small \textit{Note:} Quantile confidence intervals are obtained by bootstrap.
    \end{minipage}
\end{center}
\end{table}

\subsection{Comparison with DFJ}

Table \ref{tab:results_dfj} contrasts our estimates with \citetalias{dfj}'s SMM results. The SMM estimator uses moments consisting of median assets by groups divided by cohort and permanent income quintile in each calendar year. This makes a total of 150 moments. The cohorts are defined on a four\hyp{}year window; Cohort 1 include those who were 72\textendash6 years old in 1996, Cohort 2 are those who were 77\textendash81, Cohort 3 are those who where 82\textendash6, Cohort 4 were 87\textendash91, and Cohort 5 were 92 and older. Details are in \citetalias{dfj}.\footnote{
It is likely that conditioning on health and gender is infeasible in this setup since it would yield too many moments relative to the sample size.} The authors use the 2-step SMM estimator with optimal weighting matrix, and optimize over a wider set of parameters, which include $\beta$, and $\underline{c}$. 

Let us start by comparing the risk aversion parameter, $\nu$: 
\citetalias{dfj} obtains a much smaller point estimate than the adversarial method. This is mainly due to the estimate of the consumption floor ($\$2,665$) being much lower than the fixed value we use ($\$4,500$) in adversarial estimation. As mentioned before, a large value of risk aversion rationalizes the observed savings patterns when the consumption floor is higher. Nonetheless, \citetalias{dfj}'s risk aversion estimate increases from 3.84 to 6.04 in an alternative specification where $\underline{c}$ is fixed at \$5,000.\footnote{Their preferred specification is not the one with $\underline{c} = \$5,000$ as the authors argue that the fit of the model for this specification is not as good.}

Moving on now to the parameters governing the bequests motives,
there are no significant differences between the estimates of MPC in \citetalias{dfj} and the adversarial estimates, but there are large differences regarding the curvature of the bequests motive, $k$. Indeed, the adversarial estimates are an order of magnitude smaller than the ones from \citetalias{dfj}, which are up from around $\$10,000$ to $\$273,000$. This has important implications for the quantification of the bequest motive as a savings mechanisms: while for \citetalias{dfj} individuals only get utility from bequeathing when their assets exceed $\$36,215$, for the adversarial estimates utility is present at much smaller levels of asset and is effective already when assets exceed $\$1,320$.

In light of these results the conclusions from both estimates are different. While \citetalias{dfj} concludes that only the super rich would obtain utility from bequeathing, our estimates suggest that bequeathing matters across a much wider range of the income distribution.

Regarding the precision of the estimates, the adversarial estimates of the curvature of bequests and asset floor are significantly more precise than the ones from \citetalias{dfj}. The decrease in standard errors is in line with the theory in \cite{kmp2023adversarial} and reflects the ability of the adversarial estimator to automatically ``select'' the features of the data that are most informative about the different mechanisms, as opposed to having to specify them ex-ante as a set of moments. The gains in precision are largest when exploiting additional variation in the data in terms of gender and health.

\begin{table}[htbp]
\caption{Comparison of estimates with \citetalias{dfj}. Standard errors for the adversarial estimates are obtained by bootstrap.}
\label{tab:results_dfj}
\centering
\small
\setlength{\tabcolsep}{5pt} 
\begin{tabular}{lcccccccc}
\toprule
 & $\beta$ & $\underline{c}$ [\$] & $\nu$ & $\vartheta$ & $k$ [k\$] & MPC & $\underline{a}$ [\$] & Loss \\
\midrule
\textbf{\citetalias[Table 3]{dfj}} & 0.97 & 2,665 & 3.84 & 2,360 & 273 & 0.12 & 36,215 & $-0.67$ \\
\quad (s.e.) & (0.05) & (353) & (0.55) & (8,122) & (446) & & & \\
[3pt]
\textbf{Adversarial $X_1$} & 0.97 & 4,500 & 6.14 & 4,865 & 16.89 & 0.20 & 4,243 & $-0.67$ \\
\quad (s.e.) & & & (0.53) & (0.609$\times$10$^{6}$) & (2.19) & (0.05) & (1,293) & \\
[3pt]
\textbf{Adversarial $X_2$} & 0.97 & 4,500 & 5.99 & 1.93$\times$10$^{5}$ & 10.02 & 0.12 & 1,320 & $-0.78$ \\
\quad (s.e.) & & & (0.55) & (2.16$\times$10$^{6}$) & (3.36) & (0.05) & (1,102) & \\
\bottomrule
\end{tabular}
\end{table}

Finally, we compare the estimators in terms of the value of the loss function that they attain. The last column of Table \ref{tab:results_dfj} shows that the loss associated with the estimates in \citetalias{dfj} is comparable to the one obtained with the adversarial parameters when using $X_1$ as inputs, but is significantly higher, indicating worse fit, than the one obtained with the estimates using $X_2$. 

\subsection{Fit and Counterfactual Simulations} \label{sec:counterfactual}

We now discuss the fit of the model and also perform a few counterfactual simulations to quantify the importance of the different savings' mechanisms. All along we contrast our findings to \citetalias{dfj}.

\subsubsection{Fit}

We start by comparing the estimated asset floor to the unconditional distribution of assets one period before death.\footnote{Although the posthumous survey contains the bequest information, this may not be an accurate measure especially given that they are single individuals. Therefore, the asset one period before death can be considered a better proxy for their intended bequest.}  If individuals do not obtain utility from bequeathing above the asset floor, we should observe no bequests below that level. The asset floor implied by the adversarial estimation, $\$1,320$, is at the 22nd percentile of the distribution, while the asset floor implied by the SMM estimation, $\$36,215$, is slightly above the 40th percentile. This means that bequests, accidental or intentional, are prevalent in the sample, and a lower asset floor provides a better fit.

We next look at the ability of the model to replicate the distribution of the assets one period before death by groups of permanent income quintiles (PIqs). More specifically, we compare the sum of the assets across individuals in each group with the simulated distribution using the adversarial parameter estimates and the ones of \citetalias{dfj}. \Cref{table_beq} shows the results in lines ``DFJ baseline'' and ``Adversarial $X_2$'', respectively. Our estimates fit the assets for low PIqs well but overestimates high PIqs. On the other hand, \citetalias{dfj} estimates well the higher PIqs but underestimates the low PIqs.

\begin{figure}[htbp]
\caption{Fit by Gender and Wealth}
\centering
\includegraphics[scale=0.6]{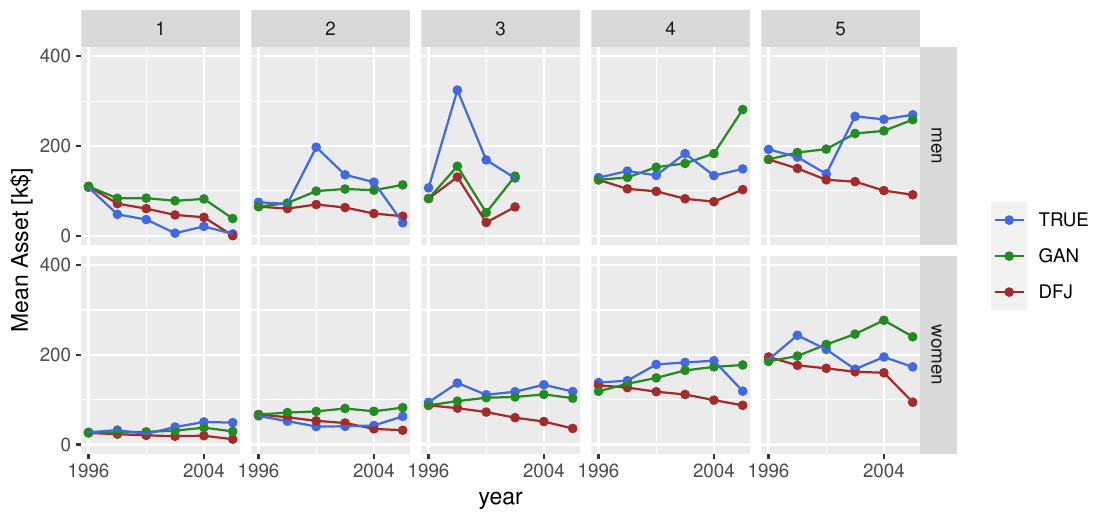}

\label{fig:fit_gender_coh2}

\vspace{0.75em} 

\begin{minipage}{0.95\textwidth}
\small \textit{Note:} Mean assets in cohort 2 separately for men and women by PIq (columns) over years. Red is DFJ, green is Adversarial $X_2$, and blue is actual data. Other cohorts exhibit similar patterns. 
\end{minipage}

\end{figure}

In Figure \ref{fig:fit_gender_coh2} we continue to examine the fit of the model, this time in terms of average asset profiles by PIqs categories separately for men and women. We are interested in assessing differences in fit by gender since this is an information that is explicitly incorporated in the adversarial estimates. Adversarial estimates seem to be closer to the sample asset profiles for women in all PI categories, except in the highest quintile, where it overestimates the accumulation of assets. Adversarial also does a good job at matching the men's profiles, especially in the highest quintiles of wealth, where survival is more likely. Additional measures of fit by wealth and cohort are in Figure \ref{fig:fit} in the Appendix.

Altogether, we conclude that the quality of the fit of the model in terms of aggregate or mean asset profiles for both sets of parameters differs across the wealth distribution. While adversarial does a better job at matching the sample in the lower to middle part of the distribution of wealth (up until the third wealth quintile), \citet{dfj} does not overshoot as much the asset dynamics at the higher quintiles. 
In terms of matching the distribution of assets one period before death, adversarial seems to be more realistically capturing smaller amounts of bequests that are left by individuals in the middle to lower part of the wealth distribution.

\begin{table}[htbp]
\caption{Fit of cumulative assets before death and counterfactual simulations}
\label{table_beq}
\centering
\small
\setlength{\tabcolsep}{6pt}
\begin{tabular}{lccccc}
\toprule
 & \multicolumn{5}{c}{\textbf{Permanent income quintile}} \\
\cmidrule(lr){2-6}
 & 1st & 2nd & 3rd & 4th & 5th \\
\midrule
\textbf{Actual data} & 18{,}191 & 25{,}266 & 42{,}006 & 50{,}495 & 85{,}814 \\
[0.5em]
\textbf{Adversarial $X_2$ baseline} & 20{,}441 & 26{,}366 & 51{,}339 & 62{,}662 & 110{,}385 \\
\quad No bequest & 17{,}644 & 21{,}587 & 42{,}586 & 50{,}631 & 95{,}212 \\
\quad (\% difference) & (13.7\%) & (18.1\%) & (17.1\%) & (19.2\%) & (13.7\%) \\
\quad No medical risk & 18{,}890 & 23{,}252 & 43{,}789 & 49{,}385 & 90{,}204 \\
\quad (\% difference) & (\hphantom{0}7.6\%) & (11.8\%) & (14.7\%) & (21.2\%) & (18.3\%) \\
[0.6em]
\textbf{\citetalias{dfj} baseline} & 16{,}527 & 19{,}672 & 38{,}157 & 42{,}737 & 83{,}814 \\
\quad No bequest & 16{,}342 & 19{,}605 & 37{,}387 & 42{,}425 & 83{,}563 \\
\quad (\% difference) & (\hphantom{0}1.1\%) & (\hphantom{0}0.3\%) & (\hphantom{0}2.1\%) & (\hphantom{0}0.7\%) & (\hphantom{0}0.5\%) \\
\quad No medical risk & 16{,}440 & 19{,}242 & 36{,}157 & 38{,}053 & 76{,}080 \\
\quad (\% difference) & (\hphantom{0}0.5\%) & (\hphantom{0}2.2\%) & (\hphantom{0}5.4\%) & (11.0\%) & (\hphantom{0}9.4\%) \\
\bottomrule
\end{tabular}
\begin{center}
     \begin{minipage}{0.95\textwidth} 
        \small \textit{Note:} ``No bequest'' rows are the simulations of the model with $\vartheta = 0$ (so $\phi \equiv 0$). ``No medical risk'' rows are the simulations of the model with $\sigma \equiv 0$ (so $\log m_t = m$). Each number is a cross\hyp{}sectional sum of assets of individuals one period before their death (in k\$), a proxy for their intended bequest. Percentages are relative to the corresponding baselines.
    \end{minipage}
\end{center}
\end{table}

\subsubsection{Counterfactual}
We perform two counterfactual simulations to measure the elderly's savings motive in terms of (i) bequest and (ii) medical expenditure risk.
We simulate the model with the same parameters except that we shut down either the bequest incentive, $\phi\equiv0$, or the medical expenditure risk, $\sigma\equiv0$.
The ``(\% difference)'' rows give the decrease of the baseline assets and counterfactual ones relative to the baseline.

The contribution of the bequest motive to the savings differs substantially between our estimates and \citetalias{dfj}.
In our estimates, the lack of the bequest motive decreases the savings by $13.7\%$ to $19.2\%$, while \citetalias{dfj} estimates suggest at most $2.1\%$ decrease.
This is largely due to the difference in the estimates of the curvature $k$.
Hence, according to the adversarial estimates, the bequest motive is an important and substantial source of savings for both the poor and the rich.


The contribution of the medical expenditure risk looks much more in line for the two models. The amount of savings to prepare for uncertain medical expenses is substantial in both predictions.
This is because rich individuals live long and hence are at high risk of large medical expenses. Poor individuals do not survive long enough to face it and are more likely to be covered by social insurance programs.

\subsection{Discussion}

Our finding that bequest motives account for roughly $13$ – $19\%$ of elderly savings aligns with the most recent literature: Studies incorporating improved medical-expense data and richer heterogeneity, such as \cite{de2025couples} and \cite{ameriks2020long}, conclude that intended bequests explain around $10$–$20\%$ percent of late-life wealth. \cite{lockwood2018incidental} similarly estimates that bequest motives account for $15–25\%$ percent of savings once annuitization choices are used to identify altruistic preferences.

Earlier work, including \cite{k2007}, \cite{kopczuk2007leave}, and \cite{dfj}, generally found smaller or more concentrated effects among the rich. By contrast, our adversarial estimation, leveraging gender- and health-driven variation in survival expectations, shows that bequest motives of comparable magnitude extend across the entire permanent-income distribution, and not only its upper tail.

\section{Conclusion}\label{sec:conc}

This paper revisits why elderly singles save by combining the \cite{dfj} life-cycle model with the adversarial structural estimator of \cite{kmp2023adversarial}. By embedding a neural-network discriminator that adaptively selects the most informative features of the data, we improve precision and identification relative to conventional SMM. The estimates imply that bequest motives explain $13$–$19\%$ percent of elderly savings, a contribution comparable in magnitude to precautionary saving against medical-expense risk. The implied asset floor, near the 22nd percentile of wealth, indicates that the desire to bequeath is not confined to the richest households. The model fits observed asset decumulation well across permanent-income quintiles and reproduces cross-sectional differences by health and gender.

These results bridge earlier findings that emphasized medical-expense risk with recent evidence highlighting heterogeneity in survival expectations. The analysis suggests that variation in health status provides a powerful source of identification for altruistic preferences in structural models of saving. 

From a methodological perspective, adversarial estimation provides a practical and efficient approach for structural models, where selecting informative moments can be difficult. Future research could extend this framework to couples and intergenerational transfers, and to general-equilibrium settings linking heterogeneity in bequest motives to aggregate saving and wealth inequality, as in \cite{de2025couples}.

\begin{singlespacing}
\bibliographystyle{ecta}
\bibliography{reference}
\end{singlespacing}

\newpage
\section*{Appendix}

\begin{figure}[htbp]
 \caption{MC approximation of Loss shape}
    \centering
    \includegraphics[width=0.7\linewidth]{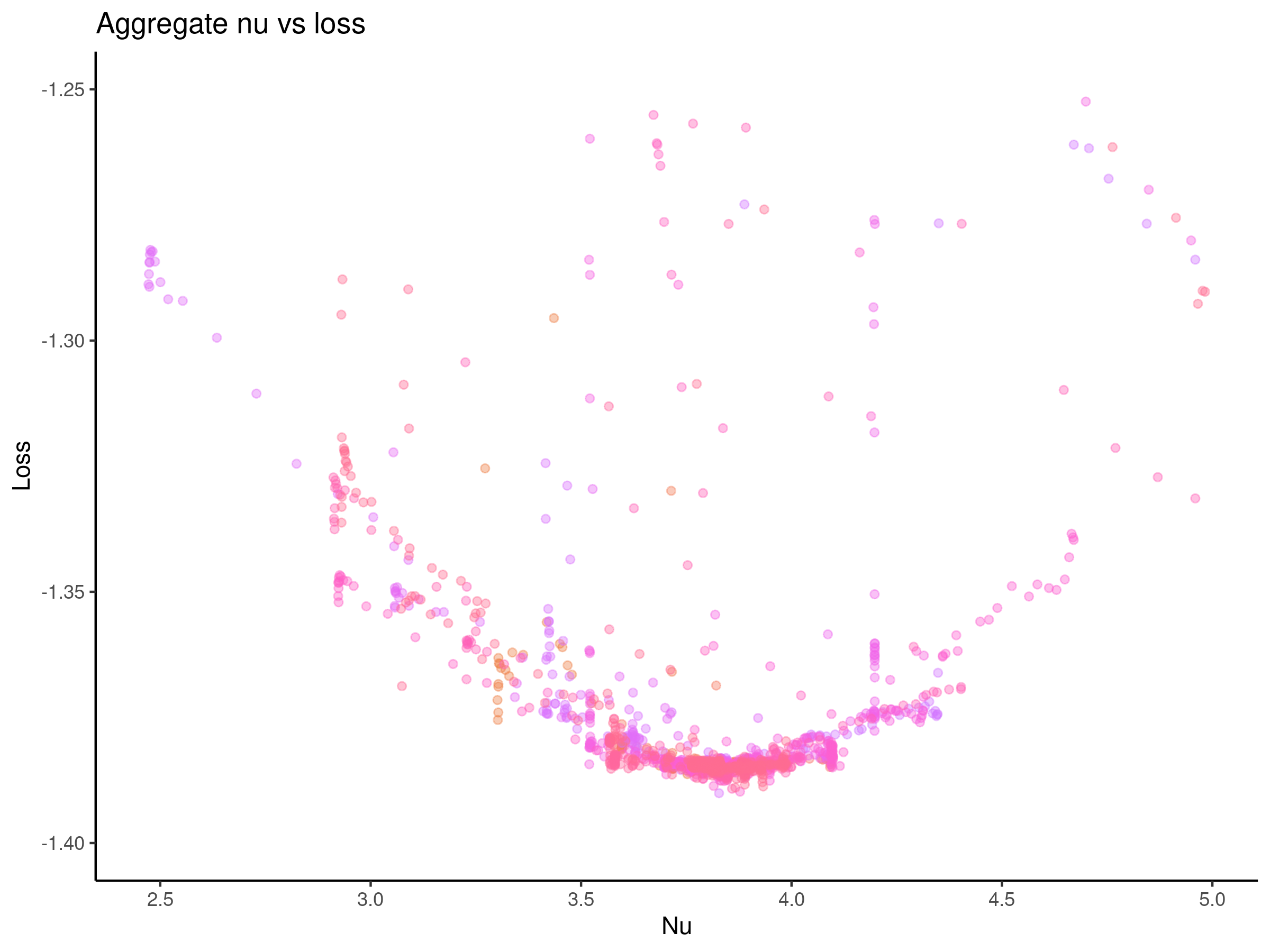}
    \label{fig:quad_form}
    \vspace{0.75em}
 \begin{minipage}{0.95\textwidth}
    \small \textit{Note:} 
The figure is constructed by pooling the parameter evolution over iterations for all experiments that uses $X_2$ as inputs and the sophisticated discriminator.
   \end{minipage}
\end{figure}

\newpage

\begin{figure}[htbp]
\centering
\caption{Fit of mean assets by cohort and wealth}
\includegraphics[scale=0.65]{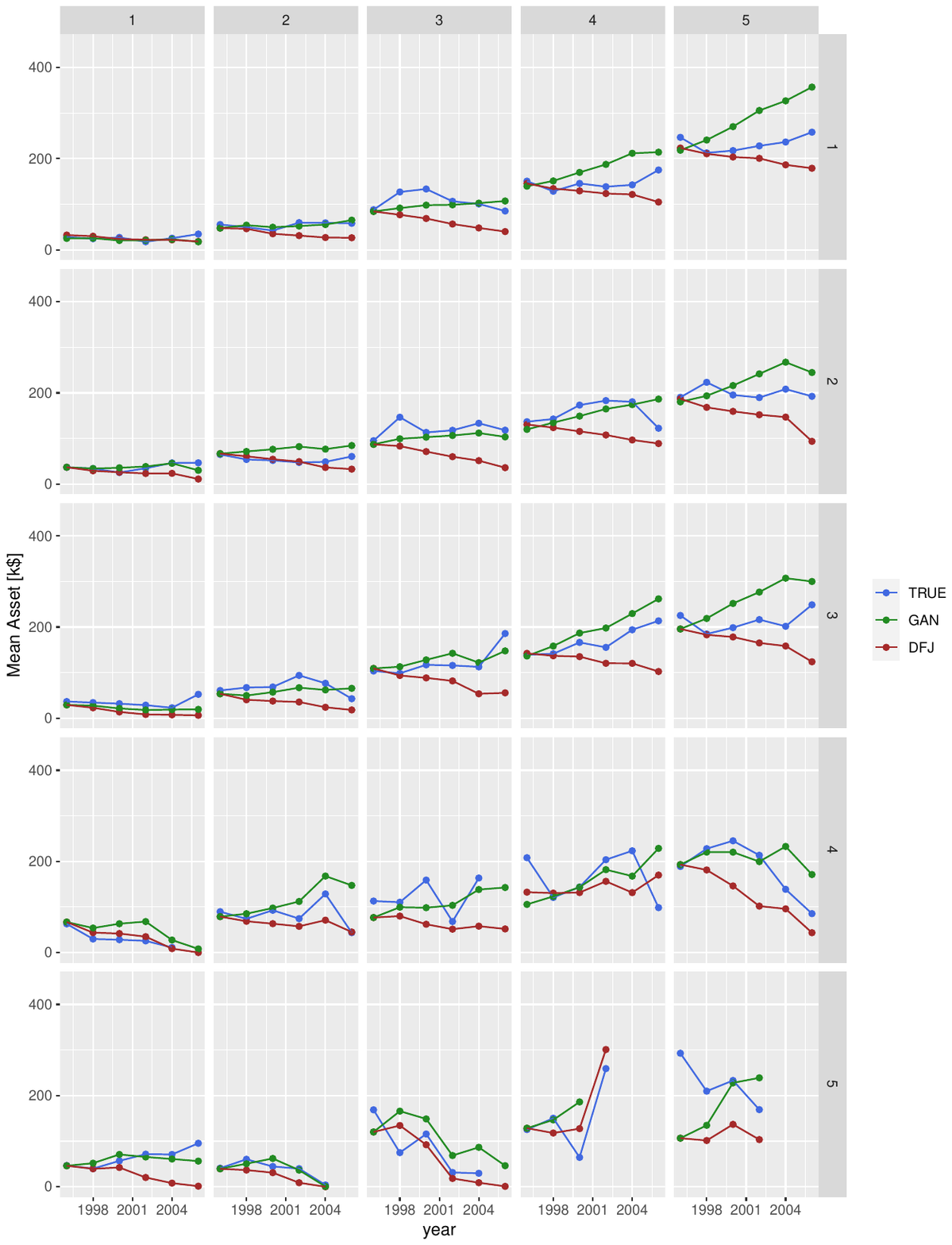}
\label{fig:fit}
\vspace{0.75em}

\begin{minipage}{0.95\textwidth}
 \small \textit{Note:}
Mean asset profiles conditional on cohort and permanent income quintiles, excluding observations above 1\% of the mean asset distribution of the actual data. Red is DFJ, green is Adversarial $X_2$, and blue is actual data. By rows we show results by different levels of PI quintiles, and by columns we show results by cohort. 
\end{minipage}
\end{figure}


\title{\MakeUppercase{WHY DO THE ELDERLY SAVE? USING HEALTH SHOCKS TO
UNCOVER BEQUESTS MOTIVES}\\{\bf Supplementary Appendix}}

\setcounter{footnote}{0}
\maketitle



\setcounter{page}{1}

\setcounter{section}{0}
\setcounter{lem}{0}
\setcounter{thm}{0}
\setcounter{exa}{0}

\renewcommand{\thesection}{S.\arabic{section}}
\renewcommand{\thelem}{S.\arabic{lem}}
\renewcommand{\thethm}{S.\arabic{thm}}
\renewcommand{\theexa}{S.\arabic{exa}}
\renewcommand{\theprop}{S.\arabic{prop}}

\renewcommand{\theHsection}{Supplement.\thesection}
\renewcommand{\theHlem}{Supplement.\thelem}
\renewcommand{\theHthm}{Supplement.\thethm}
\renewcommand{\theHexa}{Supplement.\theexa}


\section{Details on Estimation Algorithm}\label{sec:estimation_al}
In this section we discuss some common challenges in adversarial estimation. We also provide additional details on the choice of tuning parameters and implementation.

\subsection{Challenges in adversarial estimation}

The advesarial estimator in \cite{kmp2023adversarial} is inspired in the GAN estimator \cite{goodfellow2014generative}. Estimation of GAN in its original formulation (i.e. for training a generative model of images) is notoriously challenging \citep[e.g., see][]{arjovsky2017towards}. Two main issues have been raised in the literature: (i) ``mode-seeking behavior'' of the discriminator due to large differences between simulated and real sample sizes, and (ii) ``flat or vanishing gradient'' of the objective function in terms of the parameters of the structural model when simulated and actual samples are easily distinguishable by the discriminator.

Imbalances in the sample size of simulated versus actual data arise naturally in our context. Indeed, in order to reduce inflation of the variance of structural parameter estimates it is useful to choose $m >> n$. When this is the case, there is a risk that a good discriminator is one where it always predicts ``simulated data'', regardless of the input. However, this is not a useful discriminator in our endeavor. When choosing $m>n$ We follow the literature recommendation in Machine Learning and recommend  performing data augmentation on the actual samples. In particular, resampling with replacement histories of assets of individuals until both samples are even.

As per the flat gradient, we argue that this problem is not nearly as pervasive when the simulated data comes from a typical structural economic model (provided the discriminator is parsimonious enough and is not overfitting). Indeed, \cite{arjovsky2017towards} show that the problem of flat gradients is closely related to problems of overlapping support in typical generative models of images (see Lemma 1 and Theorem 2.1. in their paper), where the set of realizable images are measure zero in the space of all possible images. Typical economic models are very different from image generative models: (i) they tend to be embedded in low-dimensional spaces (the space of the endogenous outcomes),  and (ii) they tend to be parametrized by low-dimensional vectors, where searching for configurations that provide overlapping support might be computationally feasible. Nonetheless, we could still encounter this problem, especially when outcomes are discrete.

In the context of the AHEAD data, outcomes are continuous and overlapping support is not a first order problem. Nonetheless, gradients of the structural parameters tend to be close to 0 when the conditional distribution of the outcomes generated by the model and the actual data are far apart, hence making naive gradient descent a very slow strategy. We implement two speeding strategies that have recently become popular in the context of training neural networks: NAG (Nesterov Acceleredated Gradient), an accelerated gradient descent method featuring momentum \citep{nesterov27method}, and RPROP, an adaptive learning rate algorithm \citep{riedmiller1993direct}.

We now give details on our choice of tuning parameters of the algorithm for training the discriminator. 



\subsection{Computational Specifications}

We consider the set of feedforward neural network with 2-hidden layers with 20 and 10 neurons, respectively, with sigmoid activation functions in both layers, as our preferred discriminator. We rely on state of the art estimation algorithms in the R Keras package for training the discriminator. In particular, we use the default ADAM optimization algorithm, which incorporates stochastic gradient descent, and backpropagation for fast computation of gradients. For implementation of stochastic gradient descent, we select a small batch size of 120 samples per gradient calculation, and a large number of epochs (2000). As opposed to other implementations of GAN, we train the discriminator ``to completion'', and we fix the seed of the stochastic gradient to preserve non-randomness of the criterion as a function of structural parameters. We find this strategy to be the one that delivers the most reliable estimates, albeit at the cost of being computationally intensive. 

All computations were performed on the \textbf{Greene High-Performance Computing Cluster} at \textbf{New York University} using \textbf{Lenovo SD650} standard-memory nodes (16 CPU cores, 16 GB DDR4 2933 MHz RAM). Jobs were submitted through the \textbf{SLURM} workload manager using the directives
\texttt{--nodes=1 --tasks-per-node=16 --mem=4GB}, allocating 4 GB of total shared memory per node. Typical Monte Carlo runs completed in 15--24 hours with approximately 74--77\% CPU utilization.

The main program for the Monte Carlo experiments was executed in \textbf{R 4.1.2}, which calls a compiled \textbf{C executable} for data simulation. Package management was handled by \texttt{renv}. Core packages included \texttt{data.table (1.14.6)}, \texttt{Matrix (1.3--4)}, \texttt{MASS (7.3--54)}, \texttt{magrittr (2.0.3)}, \texttt{reticulate (1.36.1)}, \texttt{keras (2.15.0)}, and \texttt{tensorflow (2.15.0, CPU build)}.

Python 3.9 and TensorFlow were run through a dedicated \textbf{Conda environment}, specified in \texttt{.Renviron} (\texttt{RETICULATE\_PYTHON\_ENV=<project>/tensorflow-env}) and activated in \texttt{.Rprofile}. The C executable was compiled directly on the cluster to match system libraries:
\begin{verbatim}
gcc -O3 -DMPI_COMMENT=1 -fopenmp -o test1.exe wealth21_Singles_elena.c -lm
\end{verbatim}

A complete list of all R and Python dependencies is included in the project’s \texttt{renv.lock} file, available at:  
\url{https://github.com/elenamanresa/wealth_singlesMC/tree/parallel/clean_copy}

\defcitealias{dfj}{DFJ}

\section{Autoencoder on $X_2$} \label{sec:autoencoder}

The use of particular multilayer neural networks as discriminator can achieve faster rates of convergence than other nonparametric methods. A necessary condition, as stated in Section 4.2 in \cite{kmp2023adversarial}, is that $\log(P_0/P_\theta)$ admits the following hierarchical representation introduced in \cite{bk2019}: 

\begin{defn}[Generalized hierarchical interaction model]
Let $d \in\mathbb{N}_0$, with $d^* \in \{1,...,d\} $ and $m : \mathbbm{R}^d \rightarrow \mathbbm{R}$.
We say that $m$ admits a generalized hierarchical interaction model of order $d^*$ and level 0, if there exist $a_1,\ldots, a_{d^*} \in \mathbbm{R}^d$ and $f : \mathbbm{R}^{d^*} \rightarrow \mathbbm{R}$ such that
\[
m(x) = f(a_1' x,\ldots,a_{d^*}' x). 
\]
for all $x \in \mathbbm{R}^d$. We say that $m$ satisfies a generalized hierarchical interaction model of order $d^*$ and level $l + 1$, if there exist $K \in \mathbb{N}_0$, $g_k: \mathbbm{R}^{d^*} \rightarrow \mathbbm{R}$ and $f_{1 k},\dots,f_{d^{*}k}: \mathbbm{R}^{d} \rightarrow \mathbbm{R}$ $(k = 1,\ldots,K)$ such that $f_{1k},\ldots,f_{d^{*} k}$ $(k = 1,\dots,K)$ satisfy a generalized hierarchical model of order $l$ and 
\[
m(x) = \sum_{k=1}^K g_k(f_{1 k}(x),\dots,f_{d^{*} k}(x)) 
\]
for all $x \in \mathbbm{R}^d$.
\end{defn}

As an example, $\log(p_0/p_\theta)$ satisfies a generalized hierarchical interaction model of order $d^* = 1$ and level 0 when $p_{\theta}$ corresponds to a conditional binary choice model, such as probit or logit, irrespectively of the dimension of the conditioning covariates. 

We now provide an intuition on why fitting autoencoders on the inputs, $X_i$, can be informative of the hierarchical interaction order, $d^*$. We start by giving some background on autoencoders. 

Autoencoders are used as dimension reduction statistical models, and have been referred to as the non-linear version of PCA (e.g. see \cite{bishop2006pattern}). Autoencoders are special neural networks with three differentiated parts: encoder, bottleneck, and decoder. The encoder is typically a multilayer feedforward neural network with decreasing number of nodes in each layer. It forges a compressed representation of the inputs into the bottleneck, the hidden layer with the smallest number of nodes. The decoder takes the neurons from the bottleneck and maps it back to the output layer, increasing the number of nodes in each layer. The output layer has exactly as many nodes as the dimension of the input. Fitting an autoencoder involves minimizing the difference between the output layer and the inputs. 

Let $X \in \mathbbm{R}^d$ be a vector that can be perfectly fit into an autoencoder with $d^* < d$ neurons in the bottleneck. 
Let $X^* \in \mathbbm{R}^{d_*}$ be the output of the neurons in the bottleneck. Hence, we have: 
\[
X^*  = (en^1(\lambda_1'X), \ldots, en^{d*}(\lambda_{d^*}'X))
\]
where $en^k: \mathbbm{R}^d \rightarrow \mathbbm{R}$, with $k \in \{1,\ldots,d^*\}$ are $d$ univariate functions that map the inputs $X$ through the encoder into the $d^*$ neurons of the bottleneck. At the same time,
\[
X  = (de^1(X^*), \ldots, de^{d}(X^*))
\]
where $de^k: \mathbbm{R}^{d_*} \rightarrow \mathbbm{R}$, with $k \in \{1,\ldots,d^*\}$, are $d^*$ univariate functions that map the output of the bottleneck, $X^*$, into the $d$ neurons in the output layer  (which coincides with $X$) through the decoder.  As a result, any function of $X$, $m(X)$, can be represented as a function $g$ of $d^*$ functions of $X$. Indeed, 
\[
m(X) = m(de^1(X^*), \ldots, de^{d}(X^*)) = g(X_1^*,\ldots, X_{d^*}^*) = g(en^1(\lambda_1'X), \ldots, en^{d*}(\lambda_{d^*}'X)).
\]

Hence, $m$ admits a representation as a generalized hierarchical interaction model of some level $l$ (which depends on the exact shape on the autoencoder) and order $d^*$.

We fit autoencoders of increasing bottleneck dimension in a subset of 12 of the 21 variables in $X_2$ (excluding the constant) to investigate its underlying dimension, $d^*$. In particular, we select all binary variables: the gender indicator (1), the health status indicators over the 6 periods of observations (6), and alive/deceased indicators over the last 5 periods of observation (5). 

\begin{figure}[t!]
\caption{Fitting $X_2$ through autoencoders with increasing $d^*$.}
\centering
\includegraphics[width=0.6\textwidth]{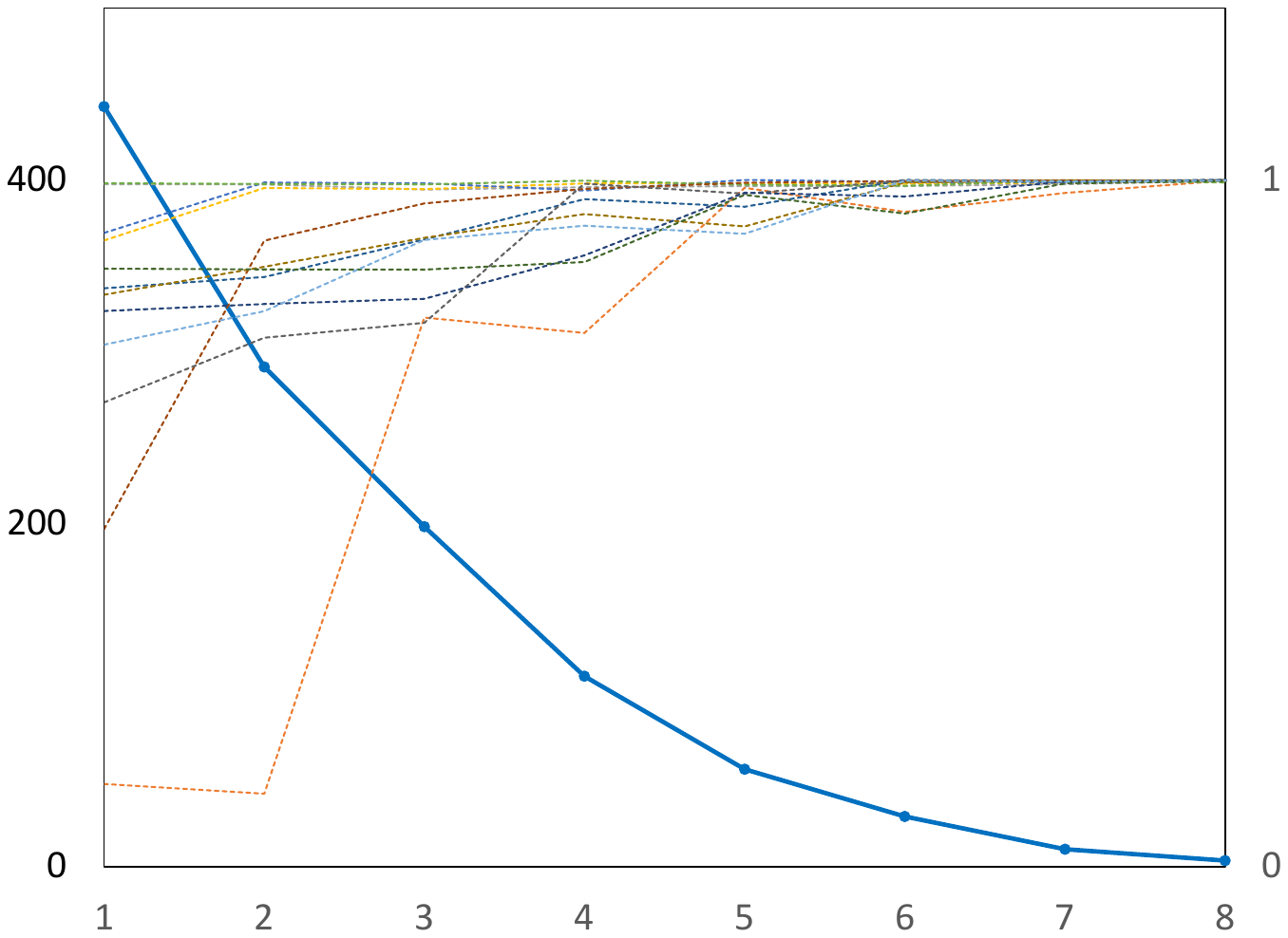}

\label{fig_autoencoders}
\vspace{0.75em}

\begin{minipage}{0.95\textwidth}
  \small \textit{Note:} The blue solid line is the MSE as a function of $d^*$. The dashed lines represent the autocorrelation of each $X_2^k$ with its prediction from the autoencoder. The left axis corresponds to MSE, the right axis corresponds to autocorrelation, and the x-axis corresponds to $d^*$.
\end{minipage}
\end{figure}

The solid blue line in Figure \ref{fig_autoencoders} represents $MSE(d^*)=\|X - \tilde{X}(d^*)\|^2$, where $\tilde{X}(d^*)$ is the output layer of an autoencoder with bottleneck size $d^*$. The remaining 12 dashed lines correspond to the correlation between the original variable $X^k$ with the prediction from the autoencoder. When $d^* = 4$, MSE has significantly reduced, and the average autocorrelation among all variables is 94.5\%. 

\begin{singlespacing}
\bibliographystyle{ecta}
\bibliography{reference}
\end{singlespacing}


\end{document}